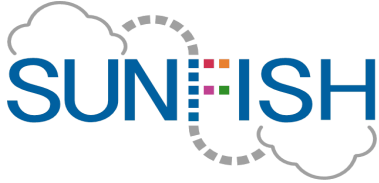
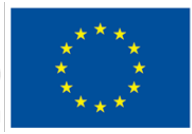

Funded by the Horizon 2020
Framework Programme of the European Union

# FaaS: Federation-as-a-Service

## The SUNFISH Cloud Federation Solution

# TECHNICAL REPORT

*Edited by*

Francesco Paolo Schiavo[1], Vladimiro Sassone[2], Luca Nicoletti[3], Andrea Margheri[2]

1 - Ministero dell'Economia e delle Finanze (Italy)
2 – University of Southampton, Electronic and Computer Science (United Kingdom)
3 – SOGEI (Italy)

SecUre iNFormatIon SHaring in federated heterogeneous private clouds (SUNFISH)
Grant Agreement Number: 644666
H2020-ICT-2014/H2020-ICT-2014-1

# Project funded by the European Commission Horizon 2020 -
The EU Framework Programme for Research and Innovation.

The SUNFISH Consortium consists of:

| Logo | Name | Country |
|---|---|---|
| | MEF – Ministero dell'Economia e delle Finanze | Italy |
| | MITA – Malta Information Technology Agency | Malta |
| | UNIROMA1 – Università degli Studi di Roma La Sapienza | Italy |
| | IBM ISRAEL – IBM Israel – Science & Technology Ltd | Israel |
| | A-SIT – Zentrum für Sichere Informationstechnologie | Austria |
| | TU GRAZ – Technische Universität Graz | Austria |
| | SOTON – University of Southampton | United Kingdom |
| | MFIN - Ministry for Finance | Malta |
| | SEROCU – South East Regional Organised Crime Unit - Cyber Crime Unit | United Kingdom |
| | CYBER – Cybernetica | Estonia |
| | PWC – PricewaterhouseCoopers Advisory SpA | Italy |

**FaaS: Federation-as-a-Service. The SUNFISH Cloud Federation Solution**

Page 4

# List of Contributors

### Chapter 1
Wayne Grixti and Jonathan Cassar from Malta Information Technology Agency

### Chapter 2
Sadek Ferdous and Mu Yang from University of Southampton

Leonardo Aniello and Gabriele Gualandi from University of Rome La Sapienza

### Chapter 3
Andreas Reiter and Bojan Suzic from Technical University of Graz

Micha Moffie from IBM Israel

### Chapter 4
Sadek Ferdous and Mu Yang from University of Southampton

Wayne Grixti and Jonathan Cassar from Malta Information Technology Agency

Andreas Reiter and Bojan Suzic from Technical University of Graz

Leonardo Aniello and Gabriele Gualandi from University of Rome La Sapienza

Baldur Kubo, Riivo Talviste, Ilja Livenson from Cybernetica

Micha Moffie, Muhammad Barham, Boris Rozenberg from IBM Israel

### Chapter 5
Leonardo Aniello and Gabriele Gualandi from University of Rome La Sapienza

**Table of contents**

**FIGURES**

# LIST OF ACRONYMS

Anonymization (ANM)

Attribute-based Access Control (ABAC)

Data Masking (DM)

Data Security (DS)

Data Transformation Service (DTS)

Federated Administration and Monitoring (FAM)

Federated Runtime Monitoring (FRM)

Federated Security Audit (FSA)

Federation-as-a-Service (FaaS)

Format Preserving Encryption (FPE)

Format Preserving Tokenisation (FPT)

Identity Management (IDM)

Infrastructure-as-a-Service (IaaS)

integration-Platform-as-a-Service (iPaaS)

Intelligent Workload Manager (IWM)

Platform-as-a-Service (PaaS)

Policy Administration Point (PAP)

Policy Decision Point (PDP)

Policy Enforcement Gateway (PEG)

Policy Information Point (PIP)

Policy Repository Point (PRP)

Proof-of-Work (PoW)

Registry Interface (RI)

Secure Multi-Party Computation (SMC)

Service Level Agreement (SLA)

Single-Sign-On (SSO)

Software-as-a-Service (SaaS)

SUNFISH Federation Agreement Contract (SFAC)

# EXECUTIVE SUMMARY

This document is the main high-level architecture specification of the SUNFISH cloud federation solution. Its main objective is to introduce the concept of Federation-as-a-Service (FaaS) and the SUNFISH platform. FaaS is the new and innovative cloud federation service proposed by the SUNFISH project. The document defines the functionalities of FaaS, its governance and precise objectives. With respect to these objectives, the document proposes the high-level architecture of the SUNFISH platform: the software architecture that permits realising a FaaS federation. More specifically, the document describes all the components forming the platform, the offered functionalities and their high-level interactions underlying the main FaaS functionalities. The document concludes by outlining the main implementation strategies towards the actual implementation of the proposed cloud federation solution.

# 1 INTRODUCTION

Cloud systems have been largely adopted in the recent years to implement any sort of computing service. The proliferation of established, reliable cloud-based system implementations has indeed lead private companies and public administrations to move their computing assets to the cloud.

Nowadays, an urgent need of companies and public administrations is to prompt and support interoperability and cooperation among the already deployed cloud systems. Indeed, it is advocated that different cloud systems aggregate themselves into homogeneous goal-oriented groups, called *cloud federations*. The underlying motivations leading to the creation of a cloud federation can be multiple: sharing of computing resources, controlled usage of third-party services or data, collaboration among entities belonging to different administrative domains Each federation aims at achieving a business need that the constituent clouds would not have achieved by themselves. A federation is thus built up around a business contract that precisely regulates the objectives and the governance of the federation.

Besides the multiple technical issues to address, creation and management of cloud federation have to face daunting security issues. Any cloud federation proposal will be adopted only if it ensures by design the security of the managed data and services. However, cloud federation is still a quite new concept that, despite the recent large research efforts, lacks of solid proposals. Therefore, the SUNFISH project aims at filling this gap by proposing a solid cloud federation platform that private and public companies can adopt. In the case of public ones, filling this gap is even more urgent as it is requested by the European Commission [1] [2].

The SUNFISH proposal consists in a new and innovative cloud federation solution, called *Federation-as-a-Service* (FaaS). It amounts to a service for clouds that enables the secure creation and management of cloud federations. This service permits federating services of any nature (i.e., infrastructural or software) and provisioning them according to access control policies dynamically enforced and monitored throughout the federation. To ensure optimal utilisation of the federated resources, FaaS offers dynamic calculation of optimal federation-based resource usage. Additionally, to further protect the security of data, FaaS features a set of data transformation services, ranging from data masking and data anonymization to secure multi-party computation, that can be used according to the needs.

A distinguishing characteristic of FaaS is its governance. To provide a cloud federation solution that any organisation can embrace, FaaS advocates a distributed and democratic governance. Indeed, the governance data is distributed along all the federation members, which enjoy the same authorities and duties on the others. To effectively enforce such a governance, FaaS proposes an innovative exploitation of blockchain, a novel technology supporting the development of distributed ledgers featuring strong integrity guarantees. A principled use of the blockchain, together with advanced integration of access control, data transformation services and runtime monitoring, allows FaaS to offer a federation service secure by-design.

In this document, we report a full-fledged presentation of FaaS, starting from its high-level functionalities and objectives, to the design and implementation strategies of its software platform, named *SUNFISH platform*. The FaaS functionalities are detailed according to a set of well-defined operating phases, which refer to both cloud administrators and cloud end-users. The design of the SUNFISH platform is first detailed by describing the functionalities of each of the components forming the platform, then by introducing the component interactions underlying the various operating phases. The feasibility of the proposed approach is finally testified by commenting on the strategies that we will pursue in the next future to actually implement a FaaS federation. Specifically, we outline the technologies and their principled orchestrations that enable automated creation and deployment of FaaS federations.

## 1.1 Setting the scene: Preliminary Definitions

This preliminary section sets the scene for the rest of this document by outlining the main entities involved in a cloud federation.

The basic ingredients of a cloud federation are the *federated services,* i.e. services offered by the individual clouds part of the federation to the other federated clouds. The service can be of any nature, from computational resources to software applications. The active part requesting a federated service, i.e. either a user or a cloud application, is called *service consumer*. Instead, the passive part providing the service when requested is called *service provider.*

The users involved in a cloud federation can be classified according to two groups: *administrators* and *service users.* It is assumed that the latter set is not directly managed by the cloud federation, which instead relies on the identity providers of the individual member clouds to which each service user belongs to. Instead, we assume that the cloud federation federates (part of) the identities of the member clouds in order to define the set of administrators of the federation. In particular, two different types of administrators are present:

1. *Member Cloud Administrator*: this role corresponds to the identity assigned to the user that owns the capabilities for administering its cloud, e.g. it can create virtual machines, access all the cloud APIs, or set Service Level Agreement (SLA) and access control for the whole cloud.
2. *Tenant Administrator:* this role corresponds to the identity assigned to the user that owns the capabilities for administering a single tenant where a service is installed, e.g. publish new service, manage virtual networks, or set SLA and access control for a specific service provided by the tenant (in compliance with the constraint posed by the Member Cloud Administrator).

## 1.2 Document Structure

The rest of this document is organised as follows. Section 2 first introduces some background concepts on cloud computing and cloud federation. Then, it presents in details the functionalities of FaaS. Section 3 details the objectives and requirements of FaaS. Section 4 reports the design of the SUNFISH platform enabling FaaS. It is first reported a high-level view of the constituent components and then their interactions to ensure the accomplishment of the FaaS functionalities. Section 5 outlines the main implementation strategies that will be pursued by the SUNFISH project to implement the proposed cloud federation approach. Section 6 concludes by reporting some final remarks on the contributions of FaaS and its further development in the following year. Finally, Appendix A reports the complete list of SUNFISH-related terms together with their definitions, while Appendix B collects all the functionalities provided by the components forming the SUNFISH platform.

# 2 CLOUD FEDERATION: BACKGROUND AND THE SUNFISH SOLUTION

In this section, we introduce the notion of FaaS envisioned and offered by the SUNFISH platform (Section 2.2). To better understand the SUNFISH proposal and its contributions, we outline a few cloud-related background concepts (Section 2.1).

## 2.1 Cloud Systems and Cloud Federations

In this section, we introduce the background concepts of cloud systems (Section 2.1.1) and of cloud federations (Section 2.1.2). We conclude by pointing out the main drawbacks of the current cloud solutions, thus highlighting the objectives for a new generation of cloud federations.

### 2.1.1 CLOUD SYSTEMS

Cloud systems can be categorised according to different aspects, ranging from the type of the delivered services, e.g. Infrastructure-as-a-Service, to the type of the architectural deployment model, e.g. Public Cloud.

The classical categorisation of cloud systems is based on the type of the delivered services, which are as follows

- *Infrastructure-as-a-Service* (IaaS): it provides cloud consumers with the capability to use cloud computing resources for deploying and running arbitrary software.

- *Platform-as-a-Service* (PaaS): it provides cloud consumers with the capability to deploy onto the cloud consumer-created or acquired applications relying on the libraries and services of the cloud platform.

- *Software-as-a-Service* (SaaS): it provides cloud consumers with the capability to use pre-defined applications running on the cloud.

A different categorisation of cloud systems is instead based on the pursued deployment model. Generally speaking, we can identify two main, opposed models: private and public cloud. Some of the main international standardisation bodies, i.e. the National Institute of standard and technology (NIST), the European Commission and the European Network and Information Security Agency (ENISA), have proposed standard definitions for private and public clouds (see, e.g., in [3] [4] [5] [6]). We summarise them in the following.

**Private cloud**. The cloud infrastructure is provisioned for the exclusive use of a single organisation. The cloud can be owned and managed by the organisation itself or by a third party (e.g. Google or Amazon). The functionalities offered by the cloud are not directly exposed to consumers outside the organisation.

**Public cloud**. The cloud infrastructure is provisioned for the general use of public users. The cloud provides different services (i.e., IaaS, PaaS or SaaS) that external users (i.e., organisations, public companies, single costumers) can buy. The cloud is deployed on the premises of the provider (e.g., Amazon or Google) and the user has no control on the actual infrastructure.

Private and public clouds are two opposed approaches to deploy cloud systems. On the one hand, a private cloud ensures full control on the computing infrastructure, workload management and service accesses, but it comes with the cost and burden of a direct management of the underlying computing infrastructure. On the other hand, a public cloud permits outsourcing the whole infrastructural management, thus reducing some internal costs and prompting flexible resource alloca-tion. However, this outsourcing prevents cloud consumers from having a tight control on the infrastructure, e.g. controlling who can access the stored data.

Private and public clouds are opposed solutions, each of them offering different characteristics and enjoying different properties. The wide adoption of cloud systems, either private or public, has indeed paved new issues to address: making already deployed clouds interconnected and cooperative. Organisations are looking for appropriate solutions to create flexible cloud systems, formed dynamically by different individual clouds that want to cooperate to achieve a business goal. In the recent years, different preliminary solutions have been proposed; we review them in the next section.

## 2.1.2 FROM CLOUD INTEROPERABILITY TO CLOUD FEDERATIONS

Many cloud computing actors have been focussing on the issue of cloud interoperability and have proposed and popularised some approaches towards this direction, see e.g. [7] [8] [9] [10] [11]. Among others, the most well-established solutions to achieve cloud interoperability are as follows:

- *Hybrid Cloud:* a cloud architecture that allows a private cloud to form a partnership with a public cloud.
- *Inter-Cloud*: technical solutions that permit different clouds to be interconnected.
- *Cloud Federation*: aggregation of different cloud systems that cooperate to achieve business goals, e.g. optimal usage of computational resources.

Even though initial comparison among these solutions are available (see, e.g., in [7] for a review), pointing out the best solution mainly depends on the application needs. In the following, we briefly comment on these solutions, while in the next section we outline the main drawbacks of the current solutions that prompted the SUNFISH project to devise its approach for cloud federation.

Hybrid clouds [3] allows an organisation that owns its private cloud to move part of its operations to external cloud providers. Single private clouds can combine their local resources with resources from a remote cloud. This solution does not involve a *service brokerage* functionality, i.e. an architectural solution that permits sharing among clouds the services they offer (besides their underlying computational resources). Currently, there are not many hybrid clouds actually in use, though some prominent initiatives have appeared on the market, e.g. by IBM and Dell.

The terms Inter-cloud and cloud federation [8] [11] are commonly referred to all the integration and aggregation solutions aiming at creating a homogeneous cloud infrastructure formed by different clouds. These solutions are in the early stage of development and several different definitions and descriptions have been used and proposed.

Inter-cloud [11] aims at creating a sort of "cloud of clouds", where each cloud is seamlessly integrated with the others thus to form a single multiple-provider infrastructure. The foundations for achieving this ambitious objective are common interfaces among all the cloud providers. Some preliminary attempts have been made (see, e.g., in [12] [13]), but some critical issues are still to be overcame before paving the way to the actual achievement of the objective. Most of all, from the perspective of commercial vendors, there is not a well-established interest in devising a single integrated multiple-provider infrastructure.

Cloud federations [8] [9] [10] consist in groups of aggregated clouds that collaborate to improve each other to achieve a business goal. Within a cloud federation the individual clouds can share resources, data and services. Different federation designs have been hypothesised [7] [9]: horizontal, i.e. cloud interconnections happen at the same service level (e.g., IaaS to IaaS), vertical, i.e. cloud interconnections happen between different levels (e.g., PaaS to IaaS). As a main ingredient, cloud federations feature a service brokerage functionality that enables the dynamic sharing of services from different providers. Besides some preliminary academic proposals, no strong cloud federation proposals are available on the market.

It clearly follows that cloud federation and Inter-cloud are relatively close at least in the objective to achieve. However, Inter-cloud is mainly based on future standards and open interfaces currently under development, while cloud federation relies on principled orchestrations of the interfaces of the cloud systems currently on the market.

Finally, we also report a few other proposals quite close to the last approaches: the *integration Platform as a Service* (iPaaS) [14] and *IaaS cloud federation platforms,* e.g. Fogbow [15]. These proposals mainly consist in tailored middleware that enables the communication of different clouds. Concerning iPaaS, it can be seen as a sort of PaaS application that permits the integration of clouds and on-premises applications. In particular, it permits creating precise data flows between the (integrated) clouds and on-premises applications. However, these flows can be neither dynamically changed nor adapted, hence its intrinsic static nature does not permit iPaaS to be exploited for advanced cloud aggregations. Instead, concerning IaaS cloud federation platforms, they propose decentralised middleware whose sole purpose is enabling and support a IaaS integration.

### 2.1.3 THE NEED FOR A NEW CLOUD FEDERATION SOLUTION

Nowadays, an urgent need of companies is to prompt and support interoperability and cooperation among the already deployed cloud systems. Among the different solutions presented in the previous section to achieve cloud interoperability, the most viable one is *cloud federation.* Indeed, hybrid clouds have been successfully exploited in various contexts, but they do not enable appropriate service brokerage and are not flexible enough to ensure, e.g., distributed control among a group of clouds; these needs are crucial in the public sector due to strict legislative requirements. Instead, the Inter-cloud solutions are still too preliminary to be considered as solid means to achieve cloud interoperability. In the case of iPaaS and the IaaS federation platforms, the main drawback is that they are specialised towards a specific functionality. Hence, the cloud federation solution seems the designed candidate to satisfy the actual needs of companies and of public administrations. However, current cloud federation proposals lack of some fundamental traits and functionalities concerning security and workload management. In the following, we identify the main ingredients for a new and innovative cloud federation solution that the SUNFISH project aims at proposing.

A new cloud federation solution must feature a service brokerage functionality that, rather than current approaches, should be able to broker any type of service (i.e., from IaaS to SaaS) in a transparent way. This brokerage activity has then to be carried out in pair with the monitoring of SLA policies that each service provider (i.e., the clouds forming the federation) agreed to ensure.

The federation must also ensure by design that the shared services are provisioned in a secure manner, thus to relieve individual clouds from additional security management. Specifically, there should be appropriate functionalities to control the accesses to services and to ensure that either service computations or data exchanges cannot cause the disclosure of confidential information. To this aim, together with an access control system distributed on all the members of the federation, a new cloud federation solution should offer secure data transformation services. A cloud willing to participate to a federation will thus enjoy the advantages of relying on advanced security-preserving functionalities.

The creation of cloud federations is commonly triggered by a business goal that the participating clouds, by cooperating together, want to achieve. Therefore, in most of the cases, it is hardly possible to identify a leader of the federation in charge of the federation management. Most of all, within specific business sectors, like the public administration one, there cannot be an organisation ruling on others. A new cloud federation solution that could be widely adopted must rely on a governance that is *distributed* and *democratic*, i.e. all federation members have the same authorities and duties. This ensures that any governance action, e.g. the enforcement of access control policies, is carried out with the consensus of all the federation members.

To meet the needs just presented, the SUNFISH project proposes an innovative cloud federation service and its underlying software platform expressly devised.

## 2.2 SUNFISH: Federation-as-a-Service (FaaS)

The SUNFISH approach to cloud federation aims at filling the gap of a security-centric design of cloud federation. The proposed approach is based on state-of-the-art design principles and technologies that enable secure sharing and management of cloud services.

The SUNFISH proposal consists of a platform that enables and facilitates the federation of individual clouds of any nature adhering to a set of pre-defined (technical) requisites (see Section 5 for further details). The proposed approach is expressly tailored to the needs for a new cloud federation solution presented in Section 2.1.3.

The SUNFISH cloud federation platform offers a service to federate clouds that is secure by design. We call this service *Federation-as-a-Service* (FaaS). FaaS permits individual clouds to federate themselves in order to share and use their computing resources, services and data. These shared entities represent the offerings of the federation that each member can use in accordance with security and SLA policies defined by single clouds and enforced by the platform.

FaaS enables the creation of cloud federations that are able to share services of any cloud level, i.e. from IaaS to PaaS and SaaS. Each of the classical cloud-based services can be registered, provisioned and controlled via the SUNFISH platform. For the sake of presentation, in the rest of this document, we use the terms "FaaS federation" and "SUNFISH federation" to refer to the cloud federation enabled by the use of the SUNFISH platform, whereas the term "service" is used to refer to any kind of cloud-based service offered by a FaaS federation (via the SUNFISH platform).

The federation approach of FaaS, differently from other proposals (see Section 2.1.3) crucially takes into account the security of the service and data. Each of the individual cloud offering a service has to provide the security and SLA policies controlling the provisioning of such a service. The effective enforcement of these policies is guaranteed by the platform. To this aim, the creation of FaaS federations requires each individual cloud to be federated not only at a service level, but also at an infrastructure level. More specifically, to participate to a FaaS federation, a cloud has to expose a set of APIs through which the SUNFISH platform can guarantee secure by-design service provisioning.

The effectiveness of FaaS federations is tightly connected to the integrity and consistency of the governance data. To store such data, FaaS relies on a specific entity, called *Registry*, that opportunely exploits blockchain-based technologies. Specifically, the use of blockchain ensures that all member clouds can control and agree on the undertaken governance actions. This approach, together with a complete distribution and appropriate replication of the SUNFISH platform components, permits ensuring a democratic and distributed cloud federation governance. Namely, each member cloud has the same authorities (i.e., the capacity of triggering or performing a governance action) over the other clouds and the same duties to carry out. Further details on this blockchain usage are presented in Section 2.2.2.2.

The design and exploitation of FaaS aims at reducing the burden needed for sharing services among individual clouds. FaaS allows individual clouds to use functionalities that they may not provide by themselves. Indeed, it fosters a shorter time-to-market development of cloud-based applications based on the creation of secure cloud federations.

To sum up, FaaS differs from the other cloud federation proposals on the following aspects: (i) it advocates a federation approach at an infrastructure level allowing the federation of any cloud ser-vice; (ii) it features a principled usage of state-of-the-art solutions for data security; (iii) it offers advanced mechanisms for enforcing and auditing security policies; (iv) it offers monitoring mechanisms to audit the enforcement of SLA policies.

In the rest of this section, we introduce the main functionalities of FaaS (Section 2.2.1) and its governance (Section 2.2.2). The description of FaaS is then completed by (i) a precise identification of objectives and requirements (Section 3); (ii) the design and usage of the SUNFISH platform (Section 4); (iii) the implementation strategies to follow to deploy the SUNFISH platform (Section 5).

## 2.2.1 THE FaaS FUNCTIONALITIES

The functionalities offered by FaaS can be grouped according to well-defined high-level operating phases. These phases and their logical relationships are shown in Figure 2. The phases concern both administrator-side operations, e.g. the creation of a cloud federation, and service consumer-side operations, e.g. the request of a service. In the following, we first detail administrator functionalities of the phases from 1 to 3 (Section 2.2.1.1), then the service consumer functionalities of the phases 4 and 5 (Section 2.2.1.2).

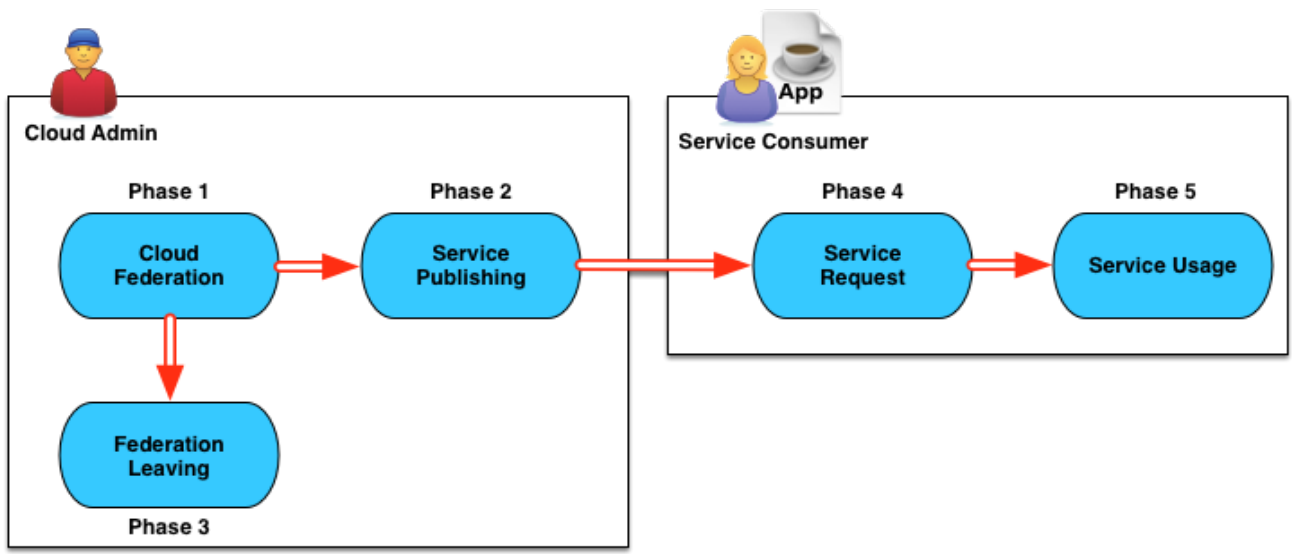

**Figure 1 The FaaS Operating Phases**

### 2.2.1.1 Functionalities for Cloud Administrators

The functionalities offered to cloud administrators by FaaS permit creating and managing a cloud federation. From a high level of abstraction, these functionalities can be grouped according to the following operating phases:

1. *Cloud Federation:* it refers to the functionalities that permit creating a cloud federation and enabling the joining of new members to an already created federation.
2. *Service Publishing:* it refers to the functionalities that permit: (i) registering a service offered by a federation member; (ii) making available the registered service to the other federation members.
3. *Federation Leaving:* it refers to the functionalities that permit a federation member to leave the federation. Leaving can be imposed when the business contract underlying the federation has been violated.

As the federation approach of FaaS relies on the federation of members at the infrastructure level, the administration functionalities mainly consist in the creation and management of logical computational spaces called *SUNFISH tenants*.

Traditionally, a *tenant* is defined as a virtual space containing computing resources exclusively assigned to a member of a (federated) cloud. A *SUNFISH tenant* is defined instead as a virtual space that is formed by resources belonging to different clouds, and can be considered as the basic building block for the federation of clouds and services. Specifically, a SUNFISH tenant can be one of the following types:

- *SUNFISH infrastructure tenant:* it represents the backbone of the FaaS federation, as it enables, on the top of it, the main functionalities underlying FaaS. There exists a single SUNFISH infrastructure tenant for each FaaS federation.
- *SUNFISH operational tenant:* it represents the provider of a service offered by a federation member. This type of tenant can be created in different forms according to the ownership of the used computing resources:
    - *standard:* the tenant is formed by computing resources owned by different members;
    - *segregated:* the tenant is formed by computing resources that are all owned by a single member.

The infrastructure tenant is the key element enabling FaaS, that is it ensures secure by-design interactions among federation members. This tenant is defined once a FaaS federation is created, whereas operational tenants are defined to publish services in a federation. The precise strategies to deploy the different tenants, as well as the constraints on the ownership of the used computational resources, are outlined in Section 5 and are the object of future project activities [16].

In the following, we further comment on tenants according to the operating phases previously introduced. For the sake of presentation, when it is clear from the context, we refer to "SUNFISH tenant" only with "tenant".

*Phase 1: Cloud Federation*

The first step to create a FaaS federation is *federating clouds.* This phase consists in federating at the infrastructure level the clouds willing to participate to a FaaS federation. To start this phase, a formal agreement among the initial member clouds has to be defined. Such an agreement, called *SUNFISH Federation Agreement Contract* (SFAC), is established within the initial FaaS governance process, whose details are reported in Section 2.2.2.

This phase can be detailed into two possible cases: (i) establishing the federation from scratch; (ii) enrolling to an already defined federation. Specifically, it means that, in the first case, the infrastructure tenant has to be created, while, in the second case, the already deployed infrastructure tenant is updated to be interconnected with the new member cloud.

The creation of the infrastructural tenant occurs between the clouds initiating a FaaS federation according to the technical details reported in Section 5. Notably, besides the initial governance, the creation of the infrastructural tenant depends on an identity federation relationship among the initial member clouds that has to be established off-line. Instead, in the case of a cloud joining an already existing federation, the following actions are performed

1. checking if the member willing to participate to a federation has the expected technical prerequisites, i.e. those requested to carry out the underlying infrastructural federation;
2. updating the infrastructure tenant to be interconnected with the new cloud;
3. update the governance data and the federation state accordingly.

Notice that the continuous availability of the requested technical prerequisites is a necessary condition for a cloud to be and remain part of a FaaS federation. Thus, according to the constituent SFAC, a member cloud can be forced to leave a federation due to the lack of these prerequisites; further details are reported as part of the phase 3.

*Phase 2: Service Publishing*

Once a FaaS federation has been created, the services offered by the member clouds have to be published in the federation. To this aim, the administrator of a cloud willing to federate a service has to follow these steps:

1. creating an operational tenant;
2. integrating as part of the newly created tenant the service to be federated;
3. registering the service to the federation by providing its SLA and access control policies.

The creation of an operational tenant is dynamically requested by a member cloud via an administration portal made available to the members. From a practical point of view, the infrastructure tenant creates a new tenant by installing a set of pre-defined components that ensure the newly created tenant to be interconnected and controlled. The created tenant can be either normal or segregated. In the latter case, a trusted environment is created where sensitive data can be managed with a high level of confidence. We call this a *segregated environment.*

Once a service has been registered, its provider can use the available administration functionalities to control and manage the service provisioning. In details, it can monitor security and performance indicators and amend the SLA and access control policies currently in force. Furthermore, concerning access control policies, it is also possible to provide administration policies that can control how the enforced access control policies are possibly modified. Indeed, according to Section 1.1, a Member Cloud Administrator can set an administration policy to which a Tenant Administrator has to adhere to for modifying the access control policies.

To ensure trustworthiness of the segregated environment, the creation of the corresponding segregated tenant adheres to specific rules about the forming computing resources: (i) they are offered by a single member cloud; (ii) they are physically located in the same data centre; (iii) their physical access is adequately regulated and controlled. Further details on these creational rules are reported in Section 5.2.2.

*Phase 3: Federation Leaving*

A member of a FaaS federation can decide in its own turn to leave a federation or, when critical violations to the SFAC have been detected, its leaving can be forced. Examples of critical violations are: absence of the technical prerequisites for carrying out the federation or continuative violations of SLA policies. There is not a pre-defined set of access control violations that lead to a federation leaving, but each SFAC can point out specific cases.

The leaving of a member is carried out according to the following steps:

1. releasing of the acquired services;
2. de-allocation of the owned tenants;
3. cleaning of all the platform data locally stored;
4. resigning from the SFAC.

These steps are carried out by the infrastructure tenant. It is also worth noticing that additional leaving duties can be required by SFAC. Thus, once a federation is created, appropriate business processes managing the member leaving must be defined as part of the infrastructure tenant.

The forced federation leaving is instead caused by the detection of member negligence in adhering to the endorsed contract, e.g. SLA or access control policies have not been fulfilled. Further details on the reasons leading to a forced federation leaving, together with those leading to a federation termination, are reported in Section 2.2.2.

### 2.2.1.2 Functionalities for Service Consumers

Once a FaaS federation has been created and its services published, service consumers (i.e., any user or application pertaining to a member cloud) can request and utilise, if authorised, the offered federated services. We comment in the following on the corresponding operating phases.

*Phase 4: Service Request*

A service consumer willing to use a service offered by a FaaS federation (i.e., any service from IaaS to SaaS) has first to submit a service request. This preparation phase has a twofold objective. On the one hand, it permits controlling the authentication and authorisation of the consumer and, on the other hand, it permits selecting the optimal service provider for such a request. What optimal means is defined according to the current state of the federation and the optimisation workload parameters chosen for the federation.

Therefore, once a service request has been submitted, its evaluation consists in the following basic steps:

1. authorising the service request according to the exposed authentication credentials;
2. calculating a list of optimal service providers that can satisfy the service request;
3. filtering the list of service providers according to the privileges of the service consumer;
4. returning the filtered list of service providers to the service consumer;
5. waiting for the choice of the service consumer and adequately setting up the needed infrastructural layer (e.g., allowing the service consumer to access the operation tenant offering the chosen service).

All the steps are carried out as result of appropriate interactions among the components of the SUNFISH platform. The final outcome of this service request phase is that the service consumer is now able to use the requested service.

*Phase 5: Service Usage*

This final phase amounts to the actual usage of the already requested, hence set up, service. During this phase, the provisioning of the service has to ensure that the corresponding access control policies are correctly enforced. Additionally, the evidences relatively to the enforcement of the corresponding SLA policies are timely gathered from the service provider. The task of identifying possible SLA violations is carried out offline with respect to the service provisioning.

The data flows generated to provision the service are indeed dynamically controlled by the access control system provided by a FaaS federation. Due to the distributed nature of a federation, the access control system is formed by distributed components deployed as part of the various tenants. These components are thus in charge of managing and controlling the data flows among service consumers and providers. To ensure that the access control decisions are correctly calculated and enforced, the FaaS federation provides a distributed system of probes that permits preventing security breaches due to integrity violations of (a part of) the access control system components.

## 2.2.2 THE FAAS GOVERNANCE

The FaaS federation service, as well as the SUNFISH platform, has been primarily conceived to fill the gaps of current cloud federation proposals, in particular relating to the management of security and governance. Concerning governance, a requirement for a new generation of cloud federation solution is avoiding centralised control. To this aim, FaaS and the SUNFISH platform have been designed to achieve a distributed and democratic control on the governance.

More specifically, a FaaS federation does not rely on a "trusted third party" acting as federation administrator. Our solution advocates parity among federation members, that is all members enjoy the same rights and duties according to the SFAC they endorsed to create the federation. Avoiding the presence of a leader ensures that all members can democratically control the governance actions carried out in the federation. The integrity and consistency of the governance data (e.g., the current access control policies in force for a specific service) are crucial to actually control the governance actions. Therefore, we foster the use of blockchain-based technologies to store and manage the governance data.

In the rest of this section, we first outline the whole governance processes underlying a FaaS federation (Section 2.2.2.1), then we introduce how the governance is supported via a principled use of blockchain-based technologies (Section 2.2.2.2).

### 2.2.2.1 An Overview of the Governance Processes

The objectives and the governance of a federation are defined according to the federation agreement contract endorsed once the federation is created. Such a contract, called *SUNFISH Federation Agreement Contract (SFAC)*, contains the business goals shared by the clouds willing to create a FaaS federation. This contract can be adapted by the federation promoters to their different needs. To support the initial governance actions, we will provide in [16] a template for SFAC that adheres to the current European legislation framework.

The business goals leading to a FaaS federation can be of different nature, spanning from technical to economical ones. By way of example, we report in the following the main goals that prompted the creation of FaaS federations within the SUNFISH use cases:

- optimised sharing of computational resources;
- enforcing advanced security controls;
- providing shorter "time-to-market" cloud applications;
- utilising advanced, secure by-design computational and data sharing services.

Besides the business goals, the agreement contract defines all the rules and data needed for the governance of the federation. The contract can be refined according to the cases, but the following data are mandatory:

- infrastructure assets (servers, computational power, storage, etc.) that are transferred by members to the federation;
- services published by members;
- (high-level) SLA policies appointed to each member.

An a priori definition of the governance rules ensures that a consensus on the behaviour of FaaS federations can be achieved between all the members.

In the following, we further comment on additional governance aspects underlying a FaaS federation.

**Governance of Federation Creation, Leaving and Termination**

As previously mentioned the creation of a new federation is prompted by specific business goals. The governance process leading to the definition of the federation agreement contract has thus to carry out an overall analysis of the business scenario, its objectives, requirements and possible threats. This analysis will result into (at least) the infrastructure assets and services to federate and the SLA contracts to adhere to.

According to the needs of a federation, the initial governance process can also prescribe specific technological solution to adopt for, e.g., ensuring service interoperability among the members. Additionally, it can also prescribe whether the federation is open, i.e. a member can join the federa-tion once it has been already created. If it is the case, when a new member joins, it has to sign the corresponding SFAC and adhere in its own turn to the reported rules.

When the business goals underlying the federation have been achieved, a member can decide to leave the federation according to the procedure previously reported. From a governance point of view, the forced leaving of a member is a more complex issue to address, as precise rules must be defined.

According to the distributed and democratic control of federations, FaaS advocates that a forced leaving can only be triggered by SLA violations and neither submitted to human decisions nor member willingness. It is thus proposed a timed alert solution. When SLA violations are detected, the corresponding member is notified and appropriate countermeasures are requested within a fixed period of time (that is defined by the SFAC). If SLA violations continue after such period of time, the member is forced to leave; the technical details to carry out this forced leaving will be reported in [16].

Finally, the FaaS governance can also support the complete termination of federation. The conditions leading to such a termination have to be properly defined in the federation agreement contract. Besides these conditions, appropriate rules to, e.g., release acquired resources and de-allocate tenants must be defined as well.

**Governance of Resources and Services**

The main objective of FaaS federations is sharing services among members. However, as the federation approach is carried out at an infrastructure level, a proper governance of computation resources has to be defined.

More specifically, it is assumed that every computing resource (part of SUNFISH tenants) is managed and made available by the administrator of the member providing such a resource. Therefore, FaaS and, consequently, the SUNFISH platform have only the duty of ensuring an optimised, secure utilisation of the federated resources.

From a service point of view, it is instead assumed that the service provider commits to the SUNFISH platform the enforcement of the access control policies it has provided when the service has been published or possibly updated via the administration functionalities. The (administrator of the) service provider is thus in charge of ensuring the correctness of such policies and it can also provide an administrative policy to regulate their management.

FaaS ensures the integrity of the committed policies and their effective enforcement. The misbehaving of service provider and consumers is dynamically monitored, e.g., according to the access attempts and the provided services. The platform cannot ensure that the services a service provider claims to offer are actually those expected, but it can control that the provider adheres to the endorsed SLA and access control policies.

To control the nature of the offered service amounts to estimate and dynamically adapt the repudiation of each service provider. Indeed, according to trustworthy feedbacks of service consumers, the provider repudiation adapts accordingly. Once a service provider experiments a low repudiation, it can be forced to leave a federation or to undertake specific countermeasures. This functionality is not currently designed as part of FaaS, but a principled exploitation of the blockchain can pave the wat to such a trustworthy repudiation system.

### 2.2.2.2 A Blockchain-based Governance

The effectiveness of the FaaS governance crucially depends on the data it relies on. Most of all, as we want to provide a distributed and democratic governance, we need to ensure that all federation members have consensus on the current data and on the undertaken governance actions. To this aim, we advocate as the cornerstone of the whole FaaS governance a blockchain-based registry for the governance data.

Blockchain is a quite novel technology that has appeared on the market in the recent years, firstly used as public ledger for the Bitcoin cryptocurrency [17]. It mainly consists of consecutive chained blocks containing records, that are replicated and stored by nodes of a p2p network. These records witness transactions occurred between some of the nodes of the network. Transactions may feature a cryptocurrency like, e.g., the Bitcoin, or other kinds of assets. The collection of transactions and their enclosing in chain blocks is carried out in a decentralised fashion by distinguished nodes of the network, called *miners*. These nodes perform a computational intensive task, called *Proof-of-Work* (PoW), which amounts to a specific hashing procedure to define new chain blocks with the consensus of all the nodes of the network. PoW allows blockchain to enjoy fascinating properties related to data integrity. Indeed, when a block is part of the chain, all the miners have agreed on its contents and, hence, it is non-repudiable and persistent. (Unless an attacker has the majority of miners' hash power, there is no computationally feasible way to alter it. However, this is not a viable attack in practice due to the size of the blockchain computational power.)

To support the FaaS governance, we envision a principled integration of (a part of) the FaaS functionalities with the blockchain-based registry. This integration permits strengthening the integrity, availability and reliability of these functionalities and, consequently, of the whole FaaS federation. Indeed, exploiting a blockchain-based registry enables trustworthy storage and retrieval of governance data. The functionalities we plan to support via this registry are as follows.

1. *Federation Contract.* The SFAC contract underlying a FaaS federation results from the initial governance activities. In order for the member clouds to control the enforcement of such a contract, the blockchain-based registry permits avoiding the use of a "trusted third-party" to store such a contract, i.e. a single-point-of-failure that might compromise the whole federation.
2. *Federated services*. To support the service brokerage functionality, the current state of the federated services has to be always available and consistent with the individual states of the members. The distributed nature of blockchain is a valuable feature that, if appropriately exploited, can ease the achievement of this objective.
3. *Access Control and SLA policies.* The access control and SLA policies in force in a federation are sensitive data to be protected to correctly provide services. To avoid attacks against their integrity, the blockchain offers highly trusted storage functionalities.
4. *Federation Monitoring*. The distributed nature of the federation requires advanced monitoring facilities that can ensure the integrity and correctness of the provisioned services. The use of blockchain allows the distributed monitoring probes to easily store the collected evidences, hence without centralising all the data and with high integrity guarantees.
5. *Data Masking.* The key ingredient of the data masking process is the storage of the masking table. To prevent integrity violation of such a table, the blockchain offers strong storage guarantees.
6. *Data Anonymization.* Traditional anonymization approaches can suffer from some typical vulnerabilities: attackers may combine previously released data and try to de-anonymize already shared data. We can use the blockchain to store and check the recipients and frequency of the sharing of anonymized data, thus to provide dynamically tuning of the anonymization parameters according to the sharing history record.

All these functionalities can be utilised via APIs (offered by a component called Registry Interface), which ensure a transparent blockchain management with respect to the FaaS operating phases.

Exploiting blockchain to support the cloud federation governance is an innovative approach fostered by the SUNFISH project. The actual implementation of this blockchain integration can be then refined according to the needs of each cloud federation application. To this aim, in the next SUNFISH activities, we will further design the integration of blockchain with the various SUNFISH platform components and, according to the use cases, the proof-of-concept implementation of these blockchain-based functionalities.

Broadly speaking, we envisage a single standalone blockchain on top of which different FaaS federations can operate, as it is graphically depicted in Figure 3. Intuitively, there will be a private blockchain to which only authorised entities, i.e. the Registry Interface components, have access. In the perspective of a European infrastructure for the Public Sector, we could envision this blockchain as a shared infrastructure between the EU members to support a new generation of public services.

It is worth mentioning that we are aware of typical disadvantages of blockchain (i.e., limited speed, limited computing resources, possible scalability issues, etc.), but some preliminary research activities undertaken within the SUNFISH project have exemplified that an appropriate use of blockchain can meet the expected requirements. Furthermore, the API-based design of the Registry Interface allows the SUNFISH platform to offer different deployment strategies for the registry in order to better adapt to different performance requirements (paying the cost of loose security guarantees). Furthre detail on the blockchain deployment will be reported in [16].

Finally, notice that to ensure the effectiveness of this blockchain approach, we must ensure the integrity of the interactions with the blockchain. To this aim, we plan to exploit trusted execution environment (e.g., TPM[1] or Intel SGX[2]) to deploy the Registry Interface component. Essentially, this trusted environment could be also exploited to deploy other crucial components of the SUNFISH platform, e.g. the components forming the access control system. [16] will report further details on the implementation and feasibility of this trusted environment, while [18] will comment on its role with respect to the whole security assurance of the SUNFISH platform.

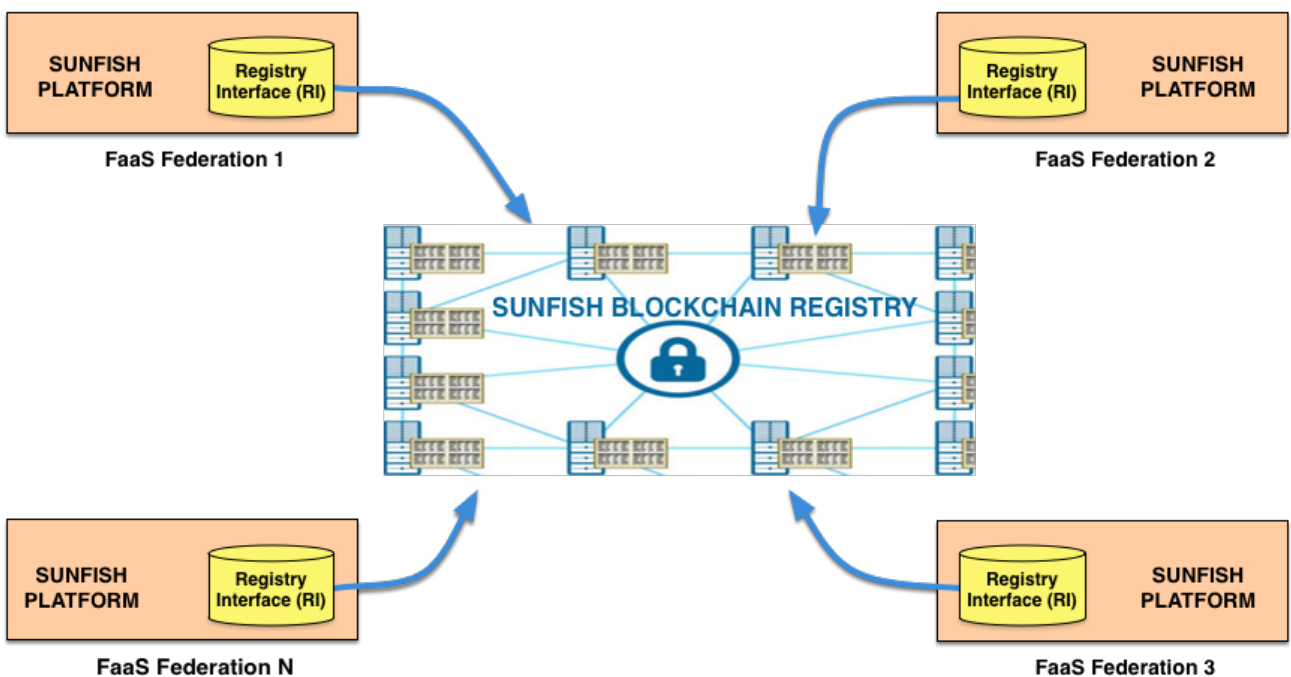

**Figure 2 Exploitation of the Blockchain-based Registry**

---

[1] Trusted Platform Module (TPM) - http://www.iso.org/iso/catalogue_detail.htm?csnumber=50970
[2] Intel SGX - https://software.intel.com/en-us/sgx

# 3 FaaS: Objectives and Requirements

The conceptualisation and design of FaaS and, consequently, of the SUNFISH platform have been prompted by a set of pre-defined objectives to achieve and requirements to guarantee. In the rest of this section, we comment on the overall objectives of FaaS (Section 3.1) and its requirement (Section 3.2).

## 3.1 Objectives

The ultimate goal of FaaS is enabling secure by-design federation of cloud systems. Indeed, a FaaS federation permits transparent and secure sharing of cloud services.

According to the current needs of cloud systems, FaaS advocates a distributed, democratic cloud federation solution. The adoption of FaaS permits reducing the management costs of cloud systems and ensuring shorter time-to-market cloud applications. Specifically, FaaS aims at achieving the following objectives:

OB-1. enabling federation of different individual clouds;

OB-2. enabling secure, controlled usage of shared cloud services;

OB-3. providing implementation strategies to deploy the proposed federation approach;

OB-4. ensuring integrity of the distributed governance data, varying from SLA and access control policies to the contract agreement;

OB-5. ensuring optimised utilisation of the computing resources of the federation;

OB-6. providing distributed infrastructure for the monitoring of SLA policies and for a reliable, monitored enforcement of access control policies;

OB-7. providing data privacy mechanisms supporting privacy regulation compliance;

OB-8. providing tight integration of secure multiparty computation with federated cloud systems.

To achieve these objectives, principled design solutions and technologies have been devised and adopted. In particular, each of the previous objectives is closely related to the outcomes of the work packages of the SUNFISH project. In the following, we thus report how these objectives will be achieved by the end of the project as result of the work packages' activities and regulated interactions. Table 1 summarises the expected results for each of the objectives.

Objectives OB-1 and OB-2 refer to the intrinsic goal of federating clouds, their services and enabling secure by-design service sharing. These objectives will be mainly achieved as result of the architectural activities of SUNFISH, i.e. this document, and the interactions with other SUNFISH activities, i.e. the documents in [19] [20] [21] [22] [23]. To measure whether OB-1 will be achieved, we set the FaaS implementation of the use cases as reference measure. Similarly, in the case of objective OB-2, we prescribe that the data security requirements provided for the use cases must be supported.

Objective OB-3 concerns the technical aspects for supporting and implementing a FaaS federation. This objective will be achieved by providing an implementation architecture document [16].

Objective OB-4 refers to the ability of FaaS of ensuring integrity of governance data. This objective will be achieved by means of a principled exploitation of blockchain technologies. This blockchain exploitation is first outlined in this document and it will be further detailed in [16].

The remaining objectives refer to specific contributions of each work packages. In details, objective OB-4 will be achieved by implementing different workload management strategies in [24]. Objective

OB-6 refers to the distributed infrastructure for SLA and access control policies, and their runtime monitoring. The runtime monitoring of the access control system permits indeed to strengthen the whole security assurance of the platform by, e.g., dealing with various threats against the integrity of access control system components.

Similarly, objective OB-7 refers to the data masking and data anonymization, and their integration with the blockchain-based data governance. The effectiveness of the last two objectives will be further addressed as part of the security assurance analysis of the SUNFISH platform [18].

Finally, objective OB-8 refers to the secure multiparty computation. This objective will be achieved by integrating the secure multiparty computation engine with the rest of the platform.

**Table 1 FaaS Objectives (where '?' refers to any number between 1 and 4)**

| N. Obj | Measure | Reference Documents |
|---|---|---|
| OB-1 | Implementation of Use Cases | [19] [20] [21] [22] [23] |
| OB-2 | Adhere to Use Cases data security requirements | [19] [20] [21] [22] [23] |
| OB-3 | Availability of an implementation architecture document | [16] |
| OB-4 | Support for blockchain-based governance | [16] |
| OB-5 | Implementation of workload management strategies | [24] |
| OB-6 | Support for the monitoring of SLA policy enforcement and for the distributed, monitored enforcement of access control policies | [19] [22] [18] |
| OB-7 | Implementation of generic data masking and anonymization services and their integration with blockchain-based governance. | [16] [18] [21] |
| OB-8 | Integration of secure multiparty computation with the platform | [21] [23] |

## 3.2 Requirements

To address the special needs of the SUNFISH context, requirements have been engineered based on an adapted SMART approach. Accordingly, requirements have been defined and formulated such that they are Specific, Measureable, Attainable, Realisable, and Traceable. This guarantees that requirements defined can serve as suitable basis for subsequent work on the targeted cloud-federation solution.

Requirements were distinguished between two types: *functional requirements* and *security requirements*. This approach has been chosen, as the methodology followed during the requirement-engineering process differs between these two types. Functional requirements are mainly derived from available SUNFISH use-case descriptions [25]. In contrast, relevant security requirements are derived from generic security requirements yielded by the SUNFISH threat model [26].

SUNFISH has defined a collection of all functional and security requirements [27], and must be seen in close relation to each other. The security requirements are intentionally kept on a more abstract level to assure a high degree of completeness, whereas the tailored security requirements can be seen as a mapping of the generic security requirements to the use cases.

Following the chosen methodologies, the SMART-based requirement-engineering process has yielded in total 16 functional requirements and 39 security requirements. All requirements have been defined on level of abstraction that considers the trade-off between completeness and richness of detail. Overlaps and relations between functional requirements and security requirements have been addressed where suitable.

# 4 THE SUNFISH CLOUD FEDERATION PLATFORM

The FaaS federation service is offered via the SUNFISH cloud federation platform, whose design and behaving is precisely defined in the rest of this section. Indeed, we first comment the overall platform architecture (Section 4.1) and its forming components (Section 0), then we define how the FaaS operating phases introduced in Section 2.2.1 are carried out according to interactions between components (Section 4.3).

## 4.1 Platform Architecture

The architecture of the SUNFISH platform is graphically described in Figure 4. The platform can be deemed as a layer of software to be opportunely deployed (see Section 5 for further details) as part of the clouds willing to participate to a FaaS federation.

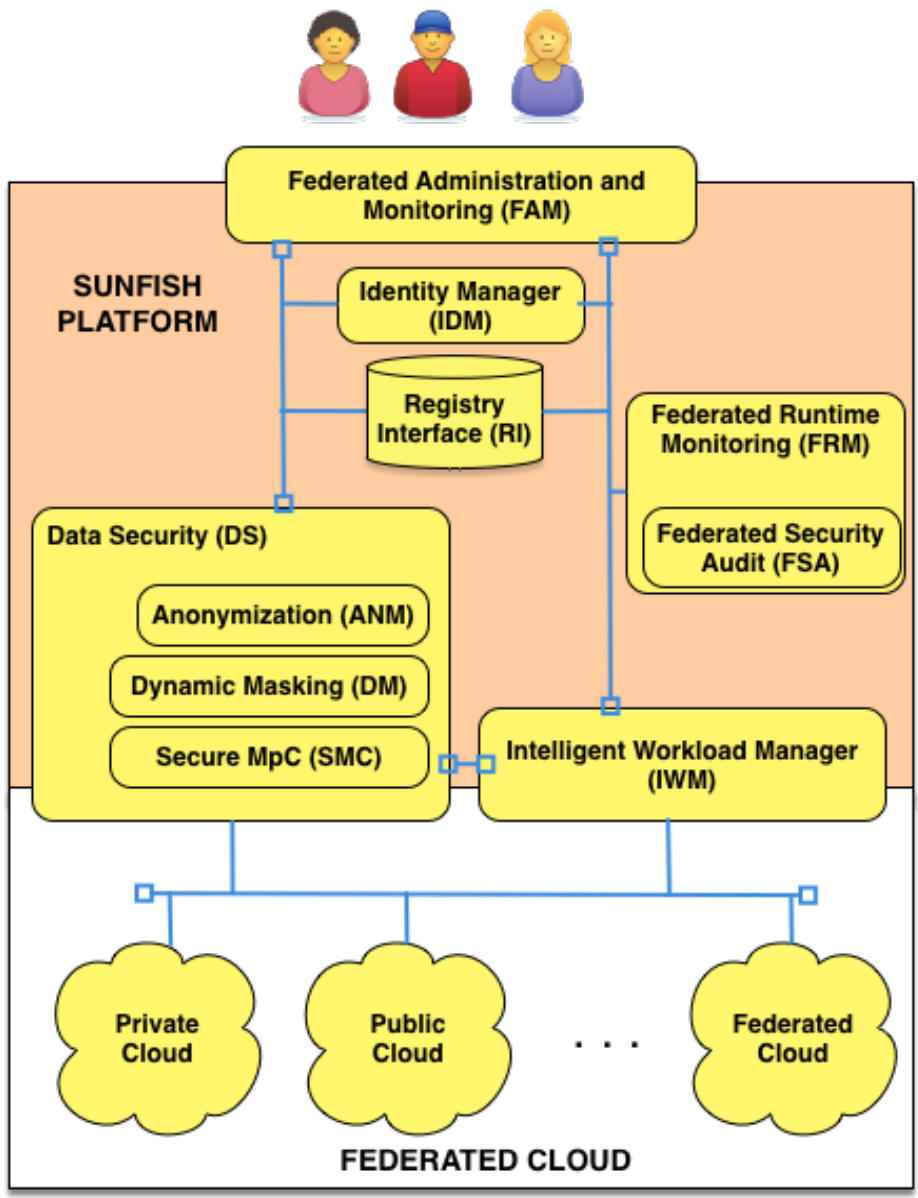

**Figure 3 The SUNFISH Platform**

The components reported in the figure provide the FaaS functionalities service and permit achieving the FaaS objectives detailed in Section 3.1. From a high-level point of view, the components are connected according to the light blue lines. The specific interactions depend on the operating phases and they are detailed in Section 4.3.

Two key components for the platform are the IDM and the RI. These components manage, respectively, the identity of all federation principals (i.e., administrators, service consumers and platform components) and the current state of the federation. It is also worth noticing that components graphically placed within others, e.g. ANM and FSA, are well-defined, standalone components that are invoked and exploited by the enclosing components.

From a low-level point of view, the federation prompted by the SUNFISH platform can be graphically depicted as in Figure 5. Indeed, the SUNFISH platform is opportunely connected via specific access and network services to the member clouds. Further details on these services are outlined in Section 5 and extensively reported in [16].

Therefore, the SUNFISH platform has been meant to be a logical software layer on the top of member clouds that, once it has been opportunely distributed, enables the creation and running of FaaS federations.

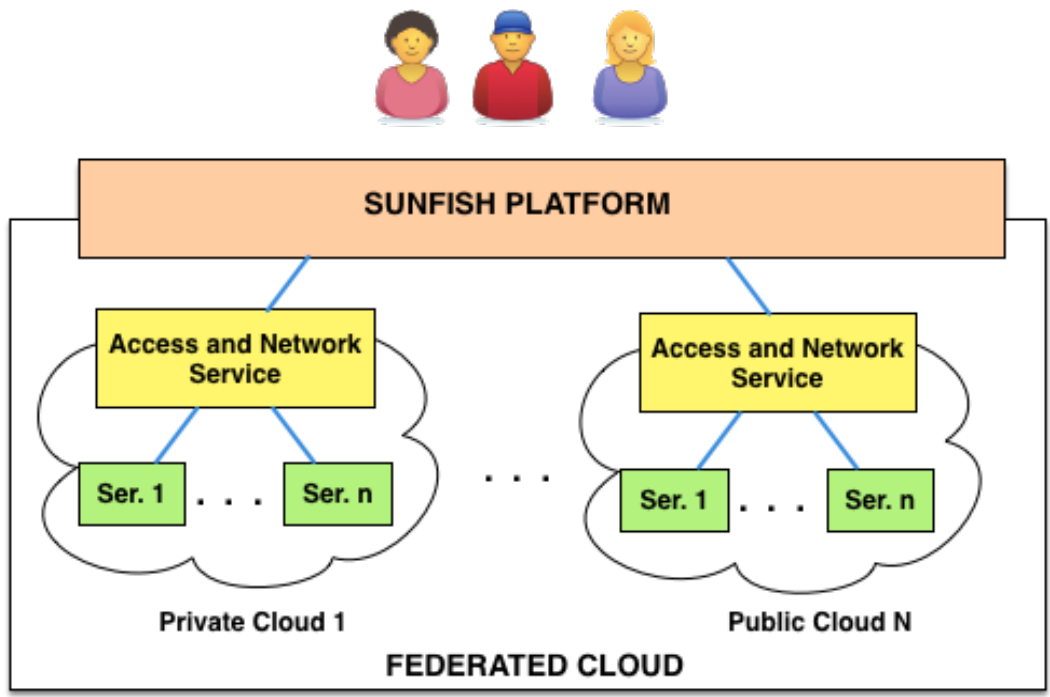

**Figure 4 A SUNFISH Federation**

## 4.2 Component Specifications

In this section, we report the high-level specification of each of the components forming the SUNFISH platform. Specifically, we report the functionalities each component will provide to the platform and hence to a FaaS federation. The full set is reported in Appendix B .

The following sections detail, respectively, each of the platform component. We refer to the corresponding specification documents for further details on component specifications.

### 4.2.1 FEDERATED ADMINISTRATION AND MONITORING

The Federated Administration and Monitoring (FAM) component represents the logical entry-points for the management of a FaaS federation, hence, for interacting with the SUNFISH platform.

The FAM provides a front-end for the administrators and service consumers of the federation based on a graphical web interface. It permits the administration of member clouds (i.e., entering and leaving a federation), tenants (i.e., creation and deploying of tenants), service publishing (i.e., registering a service to the federation), and service provisioning (i.e., management of SLA and access control policies).

Besides the administration functionalities, the FAM encompasses additional functionalities. On the one hand, it provides the reporting of possible SLA policy violations. The evidences to identify SLA violations and, consequently, to perform these reports are provided by the FRM. On the other hand, it collects and diffuses all the security alerts that the FRM and FSA components can dynamically raise and it graphically visualises these alerts by means of expressly developed dashboards.

The FAM interacts with the IWM to manage the tenants and relies on the Registry, i.e. interact with the RI, to store and load information related to the member clouds, tenants and services. The FAM also has its own database to store local configuration.

All the functionalities of the FAM can be summarised as follows.

FAM 1 Define graphical entry-points for cloud federation administration functionalities.

FAM 2 Report on SLA policy violation.

FAM 3 Collect and diffuse security alerts received by the FRM and FSA components.

FAM 4 Visualise security alerts by means of a dashboard.

FAM 5 Subscription to federated services for service consumer.

It is worth noticing that, to perform the functionality FAM 1, the FAM will rely on the communication APIs provided by the IWM to interact with the different clouds and will exploit low-level software entities that define the cloud commands to be remotely executed (see Section 5.2.1.2 for an example).

### 4.2.2 IDENTITY MANAGEMENT

The Identity Management (IDM) component consists of a set of services for authenticating the access to and within a FaaS federation. It supports the authentication of all the entities part of the federation, varying from users and administrators to service providers and platform components.

This component has been defined in its preliminary version in [28] and it mainly includes the following functional modules:

- Identity Provider: it manages the identities of the various principals of a federation.
- Authentication Server: it provides authentication credentials for each of the identities.

These modules will enable the federation of the identities belonging to the individual clouds. They thus enable the authentication of (i) administrators interacting with the FAM; (ii) service consumers interacting with IWM to request a service; (iii) service consumers interacting via the DS component with service providers; (iv) platform components interacting with the RI component. The latter functionality allows the RI to control, based on authenticating crypto-tokens, whether a component is authorised to interact with the blockchain. Furthermore, the IDM permits ensuring a single-sign-on (SSO) mechanism, i.e. a single authentication allows multiple interactions within the federation, and compliance with the eIDAS normative [29].

All the functionalities of the IDM can be summarised as follows.

IDM 1 Provide an identity to any service consumer and provider.

IDM 2 Define a SSO authentication mechanism.

IDM 3 Enable the federation of the identity managers of the individual clouds.

IDM 4 Provide endpoints for the generation of authenticating crypto-tokens.

IDM 5 Define an identity management compliant with eIDAS.

To sum up, the IDM is seamlessly integrated within the platform and is at the basis of all the flows.

### 4.2.3 REGISTRY INTERFACE

The Registry Interface (RI) is the logical component that manage the storing and retrieval operations directed to the blockchain-based registry. Recall that the Registry is a standalone entity with respect to the SUNFISH platform and is exploited by FaaS federations to store governance data.

The RI offers a set of APIs that the different components, if authorised, can invoke to store or retrieve data from the registry. The APIs support the functionalities reported in Section 2.2.2.2 and act directly on the blockchain. To control the interactions with the blockchain, the RI introduces authorisation controls based on crypto-tokens. Hence, on the base of the component identity (testified by the crypto-token), the RI decides if the component issuing the action is authorised.

All the functionalities of the RI can be summarised as follows.

RI 1 Offer a set of APIs to store and retrieve the governance data to and from the blockchain-based registry.

RI 2 Define authorisation controls on the API invocation based on crypto-tokens.

Notice also that, as discussed in Section 2.2.2.2, we plan to deploy the RI in a trusted execution environment, thus to ensure its integrity.

### 4.2.4 DATA SECURITY

The Data Security (DS) component aims at enforcing the access control policies associated with the federated services. Its main role is deciding whether to allow access requests concerning service requests (i.e., the FaaS operational phase "Service request") and provisioning (i.e., the FaaS operational phase "Service Usage").

The DS is based on Attribute-based Access Control (ABAC) [30]. The defined access control system is distributed throughout the federation and is based on a collection of logical entities whose high-level definitions follow.

- Policy Enforcement Gateway (PEG): it intercepts access requests to be authorised by managing their (secure) forwarding to the other access control system entities. Once an access decision has been calculated, it enforces such a decision in the federation.

- Policy Decision Point (PDP): it calculates the access decision for each access requests with respect to the available access control policies.
- Policy Information Point (PIP): it provides attributes representing contextual information that could be needed to calculate an access decision. It is in charge of interacting with any source of data external from the access control system.
- Policy Repository Point (PRP): it retrieves via the RI the access control policies currently in force in the federation. According to the needs, it can optimise the policy retrieval due to the incoming request to authorise.
- Policy Administration Point (PAP): it controls the administration actions on access control policies, thus deciding whether a policy amendment is authorised.

Additional details on each of these components, together with their deployment strategies, can be found in [31]. Instead, the syntax and semantics of the access control policies stored by the PRP and evaluated by the PDP are presented in [32]. Notice also that the DS is in charge of invoking, when it is needed, the so-called data transformation service (DTS) components, namely the ANM, DM and SMC components.

All the functionalities of the DS component can be summarised as follows.

DS 1 Support the evaluation and distributed enforcement of ABAC policies.

DS 2 Define invocation mechanisms for the DTS components.

DS 3 Define access controls for the operational phase "Service Request".

DS 4 Define access controls for the operational phase "Service Usage".

DS 5 Define administrative controls on the modification actions on access control policies.

### 4.2.5 SECURE MULTI-PARTY COMPUTATION

Secure Multi-Party Computation (SMC) component is one of the DTS components and provides the SMC service. SMC enables secure computation without revealing information about private inputs. It can be thus used for computing tasks on confidential data.

To carry out the SMC service, the SMC component offers the following functional modules.

- Secret sharing module: splits input data into random shares.
- Computing servers: jointly perform secure computation protocols on secret-shared data.

The SMC service is instantiated using the Sharemind technology[3], i.e. a distributed infrastructure that permits issuing controlled and encrypted communications. For each application scenario, the privacy-preserving application needs to be deployed, if pre-existing like privacy-preserving analytics or developed upon need. Once the different modules for a specific application have been defined, the SMC can be deployed as part of different SUNFISH tenants and invoked by the IWM.

All the functionalities of the SMC component can be summarised as follows.

SMC 1 Provide the secure SMC service.

SMC 2 Integrate the secure SMC service with the SUNFISH platform.

---

[3] https://sharemind.cyber.ee/technology/

## 4.2.6 DATA MASKING

The Data Masking (DM) component is one of the DTS components and provides a generic service for selectively mask personal and/or sensitive information. This service is called *masking service.* The service, given a policy and payload (e.g. text, JSON, XML), results with a masked payload. The masked payload itself is identical in format and structure to the original payload except for the personal or sensitive information which is masked.

The masking operation and its reverse operation, unmasking, are performed according to the policy provided. The policy itself allows the user to:

- select the location(s) in the payload that need to be masked (e.g. using XPath for XML documents);
- define the operation to perform on the data selected (e.g. redact, encrypt, etc.).

The DM consists of the following functional modules.

- Policy Engine: this module is able to parse and validate the DM policy.
- Masking Engine: this module is able to instantiate a specialised engine for a given policy. The instantiated engine supports the masking and unmasking processes according to the given policy.
- Masking Service: this module is responsible for managing the policy and masking engine. It implements the Rest API of the DM and manages the state (e.g. the tokenization table) that is needed to restore (unmask) previously masked payloads (in case of tokenization). It then interacts with the APIs of the RI to appropriately exploit the blockchain-based registry to store the tokenization table.

The masking engine support to following operations:

- Redact: this process will remove the selected data permanently without any way to restore it (for example replace selected text with '*****')
- Tokenization: this process will replace the selected text with a token. It may be restored by using a tokenization table managed by the Masking Service. A special case of tokenization is Format Preserving Tokenization which generates a token in a given format.
- Encryption: this process will replace the selected data with an encrypted cipher (using a key) and can be restored using the same key. It may be encrypted using the standard AES algorithm or using Format Preserving Encryption (FPE). FPE will encrypt the text into a cipher text maintaining the same format.

Further details on these masking approaches can be found in [33].

All the functionalities of the DM can be summarised as follows.

DM 1 Provide the processes for masking and unmasking of personal and sensitive data.

DM 2 Integrate the (un)masking processes with the SUNFISH platform.

DM 3 Manage the masking service state (tokenization table) via the RI.

## 4.2.7 ANONYMIZATION

The Anonymization (ANM) component is one of the DTS components and provides two complementary services:

1. Micro data anonymization: a data set (i.e., table of records) is released with the k-anonymity guaranty. This processes generalizes the QI attributes to ensure the protection of sensitive information against linkage attacks [34] using other open data sets.

2. Macro data anonymization: statistical data (i.e., summary statistics such as average, sum etc.) are released with differential privacy guaranties. This processes adds noise to the summary statistic such that the probability to identify if a single person is added or removed is extremely small.

The ANM consists of the following modules.

- K-Anonymization Engine: the anonymization engine implements the anonymization process for generating k-anonymized data.
- Differential Privacy Engine: this engine provides the core capabilities to compute the statistical data and the recommended noise to be added to the result.
- Anonymization service: this module is responsible for implementing the Rest API of the ANM module and interacting with both the K-Anonymization Engine and Differential Privacy Engine. According to the specific needs, it could then interact with the APIs of the RI to appropriately exploit the blockchain-based registry to store the anonymization history records.

The anonymization service offers the following anonymization capabilities:

- k-Anonymity: release an anonymized data set, with the guarantee that each record in a dataset is similar to at least another k-1 other records with respect to the Quasi Identifier (see [35]).
- Differential Privacy: release a noisy summary statistics while guaranteeing that individual records are protected. Effectively ensuring that adding/removing a single record to/from a database does not significantly change the outputs of statistical queries.

All the functionalities of the ANM can be summarised as follows.

ANM 1 Provide the anonymization processes of sensitive data.

ANM 2 Integrate the anonymization process with the SUNFISH platform.

### 4.2.8 INTELLIGENT WORKLOAD MANAGER

The Intelligent Workload Manager (IWM) is a component acting as the service broker of the federation. Its main duty is to perform service brokerage of the federated services offered by the member clouds. In particular, once a service consumer requests a service, it provides an optimal federation-based workload deployment target to satisfy such a service request. To calculate it, the IWM retrieves via the RI the current state of the platform, i.e. the set of federated services and their meta information (e.g., the available capacity for a certain service), and resolves an optimisation problem. To actually deploy the workload, the IWM interacts directly with the clouds to, e.g., create / delete virtual machines running on the federated clouds.

The IWM can provide different workload management strategies optimised according to different parameters, e.g. cost or actual availability SLAs. It thus solves an optimization problem based on the current state of the federation and the requested service.

The IWM consist of the following modules.

- Validator: it validates and updates the current state of a SUNFISH federation. It also interacts with the PEG to check the deployment plan against current access control policies.
- Optimiser: it calculates the matching targets for the workload based on the incoming service request and targets permitted by the DS.
- Executor: it implements the business logic abstraction of cloud operations, e.g. VM provisioning or VPC creation.
- Adapter: it permits interacting with the individual clouds to invoke their APIs. These functionalities are exploited to apply the calculated workload strategy and by the FAM to admin-ister the individual clouds. The adapter is specialised to interact with state-of-the-art cloud systems, e.g. OpenStack, AWS, Azure.

Additional details on these modules and their interactions are reported in [24].

All the functionalities of the IWM can be summarised as follows.

IWM 1 Provide the computation of optimised federation-based workload deployment targets.

IWM 2 Support different optimisation parameters for the calculation of the workload deployment targets.

IWM 3 Provide support for interaction with APIs of the federated clouds, both private and public.

IWM 4 Integrate DS policies with the optimisation path to filter out requestor-specific targets.

### 4.2.9 FEDERATED RUNTIME MONITORING

The Federated Runtime Monitoring (FRM) component provides a distributed infrastructure to intercept (via transparent plug-in proxies) every access control request received and possibly authorised by the DS. For each intercepted interaction, metadata of interest are extracted and validated with respect to the access control policies currently in force. If policy violations are detected, the FRM raises a security alert to the FAM. The FRM relies on RI to store the access control logs and to retrieve the current access control policies. Additionally, the FRM provides access logs (based on the gathered access data) to the FSA for performing additional reasoning tasks to possibly discover anomalies or security breaches.

The FRM consists of the following modules.

- Smart Agents: they correspond to the probes distributed on the access control system to collect the data on the authorisation of access requests.
- Policy Violation Engine: it validates the data gathered by the smart agents with respect to the current access control policies.

The preliminary specification of these modules is reported in [36].

All the functionalities of the FRM can be summarised as follows.

FRM 1 Provide distributed probes to monitor the access control system.

FRM 2 Detect access control policy violations by analysing the collected access control data.

FRM 3 Raise alerts to the FAM to signal access control violations.

FRM 4 Provide to the FSA the access logs to perform its reasoning tasks.

### 4.2.10 FEDERATED SECURITY AUDIT

The Federated Security Audit (FSA) component provides an automatic detection service against vulnerabilities in the distributed access control system and against security breaches possibly occurred within the federation. The FSA exploits machine learning techniques to figure out the high-level activities occurred in the federation and, consequently, to identify vulnerabilities and security breaches. When vulnerabilities or security breaches are detected, the FSA raises the corresponding alert to the FAM.

The FSA consists of a single module that, given in input the access control logs provided by the FRM, detects vulnerabilities and security breaches by applying the following techniques:

- Data Aggregation: the ability to aggregate low-level database activities into a higher-level business transaction will make activities easier to monitor and audit. Moreover, "out-of-context" activities may point to suspicious behaviours, which in turn could indicate an attempt for attack.
- Role Mining: the goal here is to learn roles from actual usage information and to employ the learned roles for the purpose of vulnerabilities identification and for the purpose of anomaly detection.

These techniques are further detailed in [37].

All the functionalities of the FSA can be summarised as follows.

FSA 1 Detect vulnerabilities in existing access control mechanisms by analysing access logs.

FSA 2 Detect security breaches by analysing access logs.

FSA 3 Rise alerts to the FAM to signal vulnerabilities and security breaches.

## 4.3 Component Interactions

In this section, we detail the interactions among the SUNFISH platform components that occur to carry out each of the FaaS operating phases presented in Section 2.2.1. In the rest of this section, we present and comment on these interactions by means of some sequence diagrams.

The diagrams are complaint with the component functionalities just presented. It is also worth noticing that these diagrams represent the logical interactions between components, not the low-level means actually used to carry out these interactions; the latter will be reported in [16]. Furthermore, notice that the diagrams have been designed with a security-centric approach and by pursuing a principled exploitation of all the data security-related components of SUNFISH. However, they will be specialised in [16] as result of the activities relating to [18].

### 4.3.1 PHASE 1: CLOUD FEDERATION

The operating phase 1 refers to the federation of a cloud with an already existing FaaS federations. The interactions of the involved platform components are graphically reported in Figure 6.

When this phase takes place, it is assumed that an initial setting up phase has been already carried out by at least two clouds that wanted to create a FaaS federation. Such an initial phase consists, on the one hand, in the governance process described in Section 2.2.2 and, on the other hand, in the creation and deployment of the infrastructure tenant underlying the federation presented in Section 5.2.1. The initial governance process does not require interactions among components of the SUNFISH platform, but it deploys them to form the infrastructure tenant and creating a FaaS federation. Hence, we do not report here a specific sequence diagram.

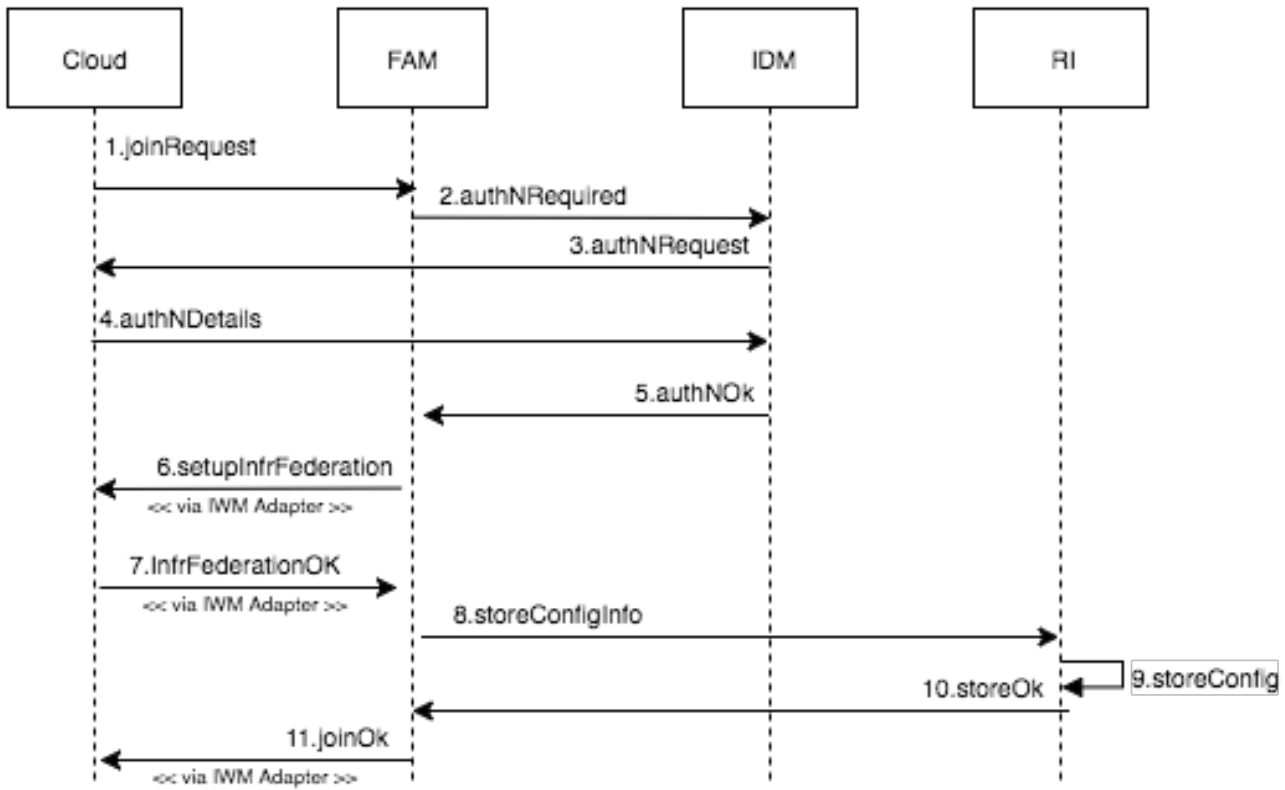

**Figure 5 Sequence Diagram Phase 1**

A cloud that wants to become a member of a FaaS federation triggers the starting of this phase by sending a specific request to the designed federation. The consequent interactions between the involved platform components are commented in the following.

- Step 1: the (administrator of the) cloud asks to the FAM of the chosen federation to become one of its member. For the sake of presentation, it is here assumed that the governance process of signing the SFAC has been already completed.

- Steps 2-5: the FAM requires to the IDM to authenticate the candidate member cloud. To this aim, the IDM in its own turn interacts with the cloud to check its authentication credentials and provide them to the FAM (see [28] for further details).

- Steps 6-7: the FAM, via the APIs of the IWM Adapter, opportunely federates the infrastructure of the cloud candidate (see Section 5 for further details).

- Steps 8-10: the FAM stores via the RI all the needed configuration data. Besides the infrastructure parameters like, e.g., the network configuration, it is stored a non repudiable evidence of the enrolment of the cloud to the federation. This evidence is thus a reference to the SFAC signed during the governance process.

- Step 11: the FAM notifies, via the IWM Adapter, the successful enrolment of the cloud to the federation.

It is worth noticing that if any of the previous interactions cannot be successfully completed (e.g., the authentication fails or the cloud does not expose enough APIs to set the federation infrastructure), the whole phase is aborted and the cloud is not accepted as a federation member. Furthermore, steps 3-6 are feasible due to the offline governance phase, i.e. it is assumed that an initial federation of the IDMs of the single clouds has been already defined. As we further detail in Section 5.2.1, once a federation has been settled, the IDM completely federates the individual cloud IDMs by being compliant with the eIDAS regulatory environment [29].

## 4.3.2 PHASE 2: SERVICE PUBLISHING

When a cloud is participating in a FaaS federation, hence the phase 1 has already successfully completed, the (administrator of the) cloud can federate its services by opportunely publishing them on the platform. The interaction between the involved platform components are graphically reported in Figure 7 and commented as follows.

- Step 1: the (administrator of the) cloud asks to publish one of its services to the FAM and, together with the identifying credentials of the service, it provides to the FAM the access control and SLA policies referring to the service usage (the policies can be changed afterwards via the functionalities offered by the FAM).

- Steps 2-5: as in the case of phase 1, the FAM requires to the IDM to authenticate the cloud issuing t the publishing request.

- Step 6: the FAM executes a sanity check on the received policies. The checks may vary from syntax controls (e.g., to assess the compliance with the XACML standard), to semantics controls possibly prescribed by the agreement contract (e.g., to assess the nature of the service levels of the SLA policies). Concerning access control policies, the check will rely on APIs offered by the DS.

- Steps 7-9: the FAM stores via the RI the identification credentials of the federated service and its access control and SLA policies.

- Step 10: the FAM, via the APIs of the IWM Adapter, finalises the publishing of the service and notifies the member cloud that the service is now part of the FaaS federation.

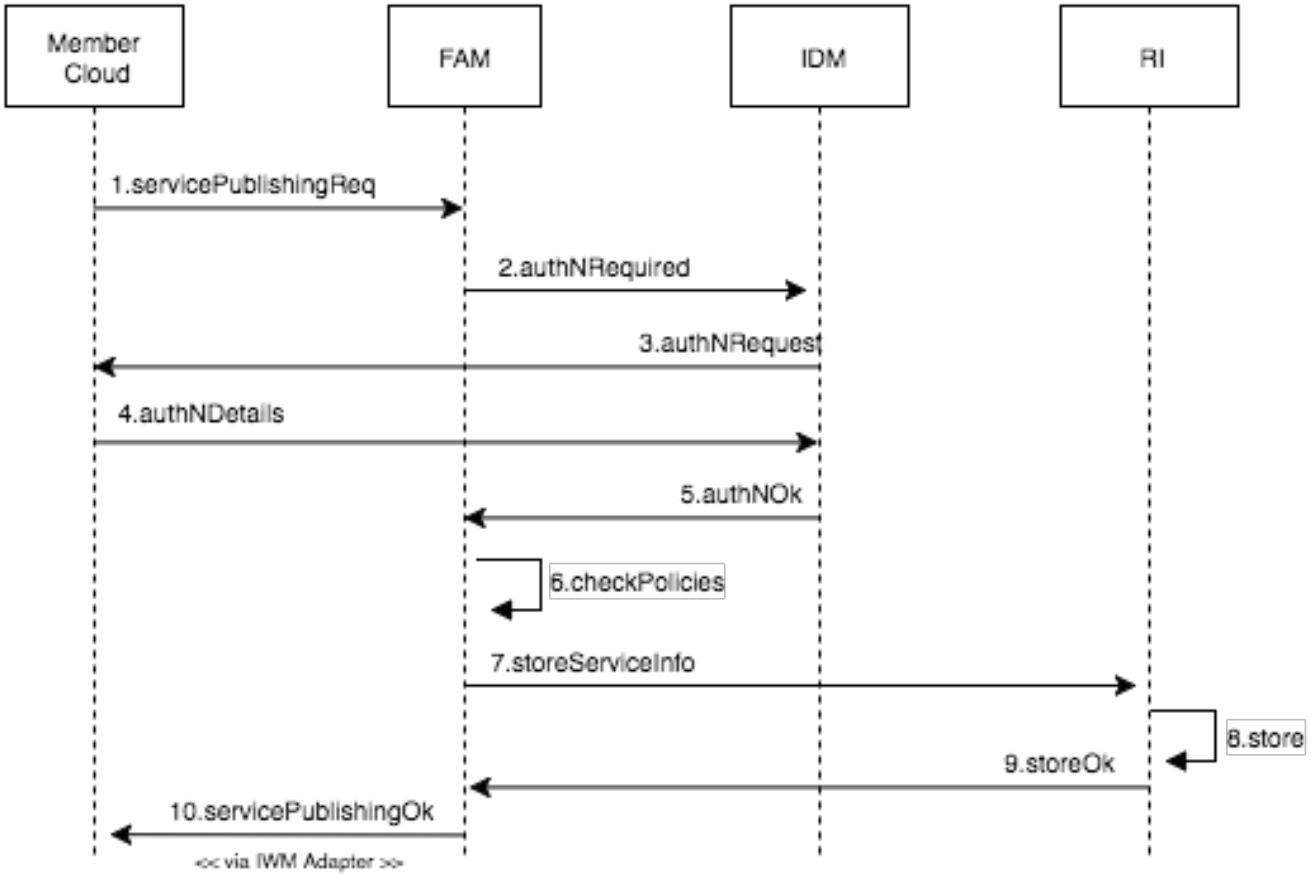

**Figure 6 Sequence Diagram Phase 2**

As in the previous phase, if any of the interactions cannot successfully complete, the service is not published on the platform and hence not federated.

### 4.3.3 PHASE 3: FEDERATION LEAVING

A cloud participating in a FaaS federation can decide to leave the federation. Indeed, the (administrator of the) member cloud triggers the starting of this phase by sending a specific request to the FAM of the federation. The corresponding sequence of steps is graphically reported Figure 8. In the case of a forced federation leaving, the FAM carries out the same steps in its own turn without being triggered by the member cloud (i.e., it directly starts from step 6).

The corresponding interactions between the platform components are commented as follows.

- Step 1: the (administrator of the) member cloud asks the FAM to leave the federation.

- Steps 2-5: as in the previous phases, the FAM requires to the IDM to authenticate the member cloud submitting the leaving request.

- Step 6: the FAM opportunely interacts, via the APIs of the IWM Adapter, with the member cloud to update the infrastructure tenant and deallocate all the services that the cloud may possibly own in the federation. These interactions rely on the API exposed by the member cloud (see Section 5).

- Steps 7-9: the FAM removes via the RI the configuration data referring to the leaving member and its federated services.

- Step 10: the FAM notifies, via the IWM Adapter, the leaving to the member cloud.

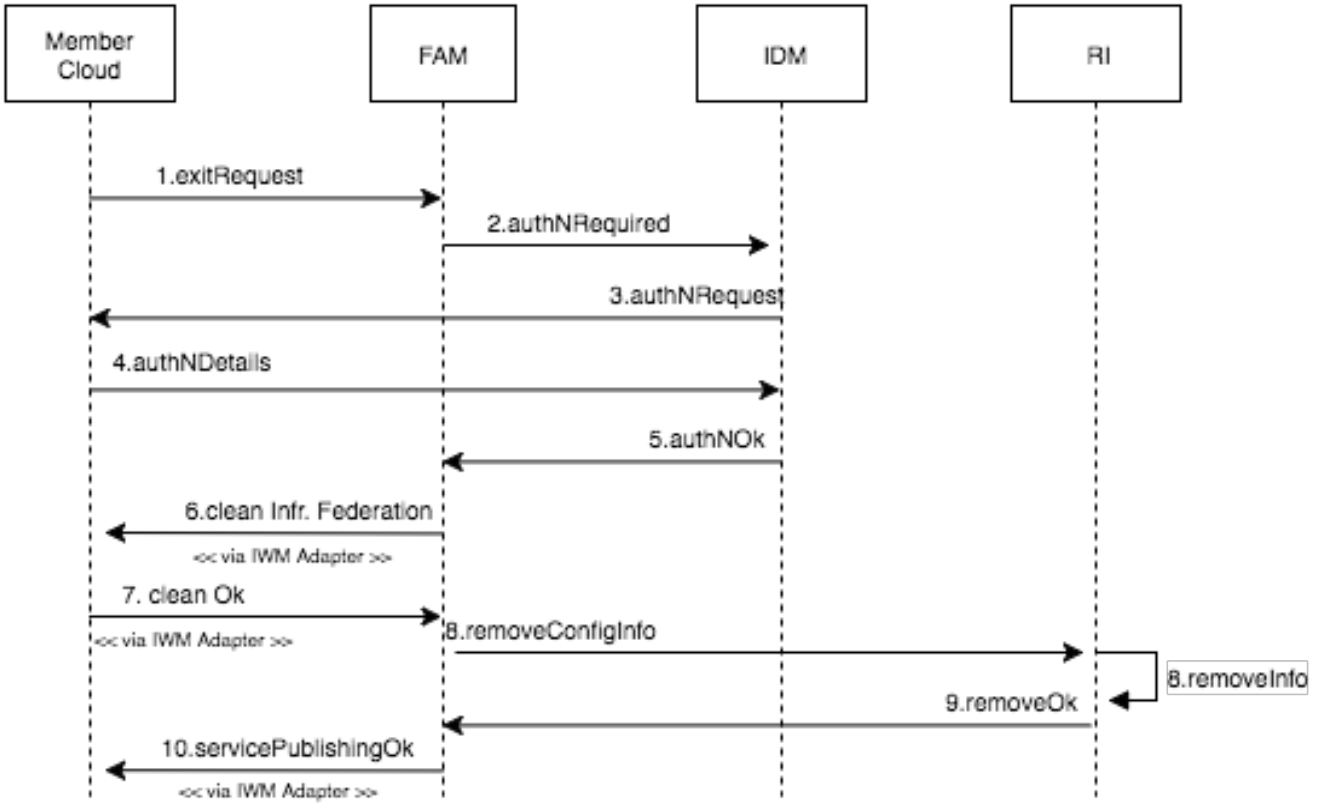

**Figure 7 Sequence Diagram Phase 3**

As mentioned in Section 2.2.2, each SFAC can prescribe specific leaving procedures. Therefore, during the step 6 additional prescribed procedures for, e.g., cleaning possible sensible data can be carried out.

### 4.3.4 PHASE 4: SERVICE REQUEST

The phase 4, as well as the phase 5, has as main actor a service consumer (i.e., a user or an application) willing to request and hence use a federated service offered by a FaaS federation. To this aim, a service consumer asks to the IWM for the available services matching with some provided characteristics. The consequent interactions among the involved platform components are graphically depicted in Figure 9 and commented as follows.

- Steps 1-3: the IWM by means of the IDM authenticates the service consumer. Differently from previous phases, we use simplified interactions with the IDM because this authentication is intended to be carried out at a lower level, e.g. at the level of the network via firewall technologies. This design choice depends on the communication means used to communicate between cloud members and the IWM; some preliminary details are reported in Section 5.1.2.

- Step 4: the service consumer asks the IWM for a service. Together with the request, the service consumer provides a set of characteristics to identify the requested service.

- Steps 5-6: the IWM retrieves via the RI the information on the available services that match the provided characteristics.

- Steps 7-8: the IWM asks to the DS, for each service in the list, whether the service consumer is authorised to see and hence use the retrieved services.

- Step 10: the IWM, on the basis of the authorisation decisions made by the DS, calculates an optimal list of services offered by the federation among which the service consumer can choose.
- Step 11: the optimal filtered list is sent by the IWM to the service consumer.
- Step 12: the service consumer requests IWM to the chosen provider.

The service consumer can now choose a service provider among those in the received list and then interact with it according to the phase 5.

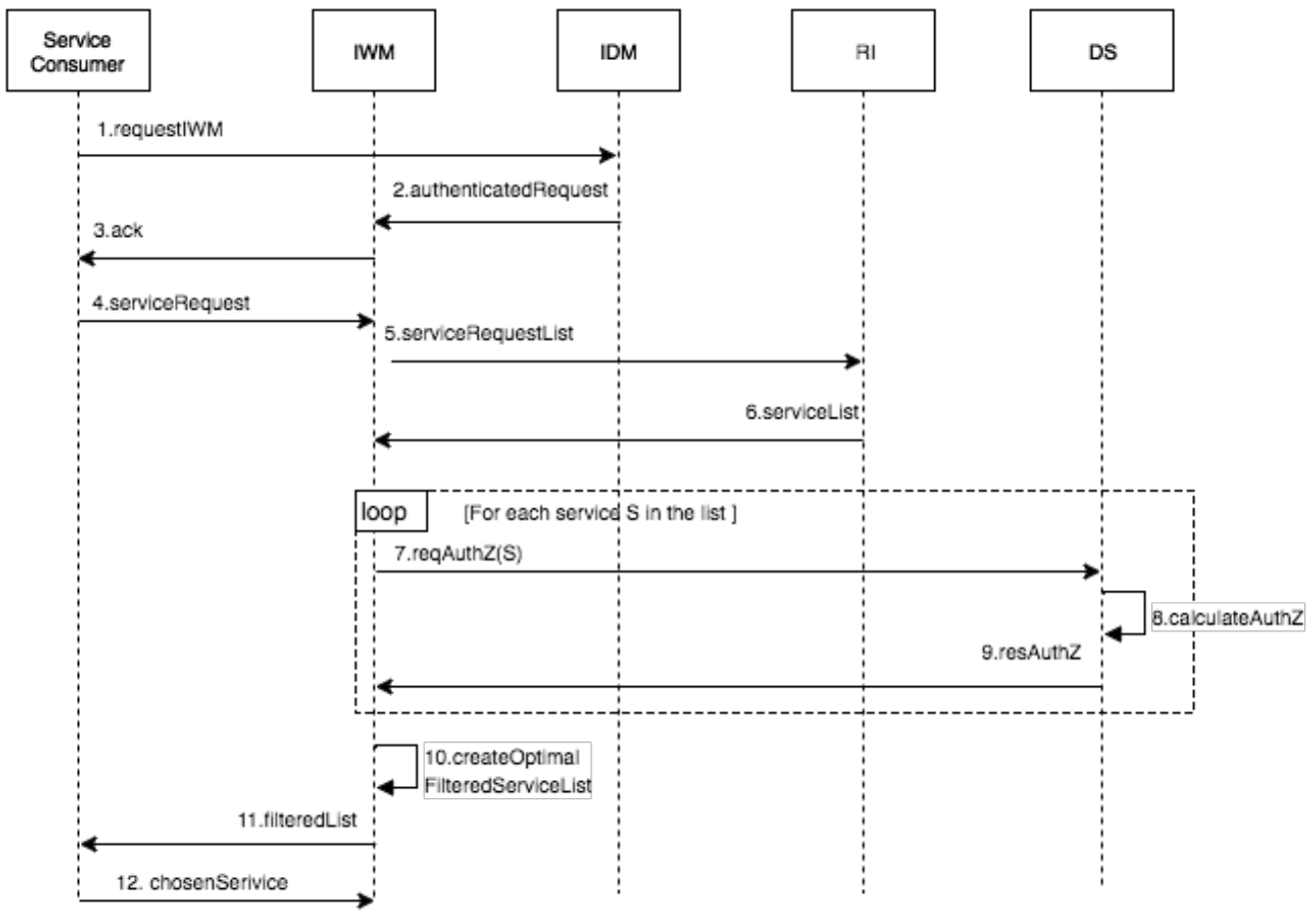

**Figure 8 Sequence Diagram Phase 4**

### 4.3.5 PHASE 5: SERVICE USAGE

In this last phase, the service consumer is interacting with the service provider chosen as result of the phase 4. This phase shows how the provisioning of a federated service is controlled and monitored by the SUNFISH platform. The interactions among the involved platform components are graphically depicted in Figure 10 and commented as follows.

- Step 1: the service consumer, which is assumed already authenticated (i.e., it exposes an authentication evidence), is using a service and the corresponding request to the service provider is transparently submitted to the DS for the enforcement of the access control policies.
- Step 2: the DS checks (by possibly interacting with the IDM) the authentication evidence of the service consumer.

- Step 3: the DS calculates the access control decision for the received service usage request; see [32] and [31] for further details.

- Steps 4-6: the FRM monitors the evaluation and enforcement of the access control policies by relying on the RI to store the logs.

- Step 7: the DS enforces the calculated access control decision[4]. We assume in the diagram that the access request has been permitted.

- Steps 8-9: the service provider provides the requested service by returning a certain result. According to the cases, the DS can invoke on the result a DTS component.

- Step 10: the secured result is sent to the service consumer.

For the sake of presentation, we represent in the diagram all the DTS in the same manner. However, the SMC is exploited to provide a service and is not applied on the result a service calculates. Therefore, the SMC is actually part of the step 8 and not of the step 9.

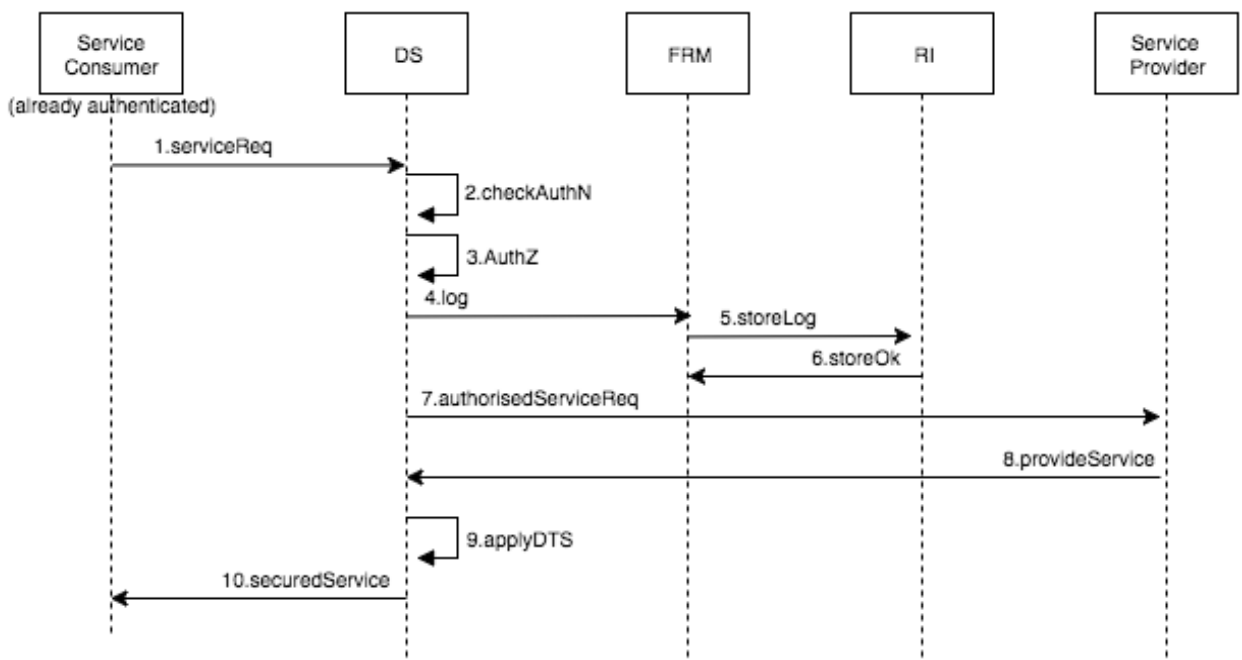

**Figure 9 Sequence Diagram Phase 5**

---

[4] Again, this step is monitored by the FRM, but for the sake of presentation we do not explicitly report the corresponding interactions between the DS and FRM components. All the details are anyway available in [35].

# 5 IMPLEMENTATION STRATEGIES

In this section, we outline the main implementation strategies that will be mainly object of [16].

The rest of this section is organised as follows. We first give a glance at the whole implementation strategies of a FaaS federation and the needed technical prerequisites (Section 5.1). Then, we present the automated strategies pursue to create and deploy tenants (Section 5.2).

## 5.1 An Overview

The key logical ingredients to implement the SUNFISH platform, as well as the deployment of a FaaS federation, are the SUNFISH tenants. The different types of tenants are introduced in Section 2.2.1, here we detail how they can be created on physical cloud machines (Section 5.1.1) and how deployed tenants can communicate among themselves (Section 5.1.2).

It is worth noticing that the functionalities presented in the following sections can be actually deployed only if every cloud adheres to some specific technical prerequisites. They indeed allow the federation at the infrastructural level to exist. The complete, detailed list of prerequisites will be reported in [16], a high-level description of them is summarised as follows.

- *Cloud resource management*: it is required that the SUNFISH platform is able to interact with the manager of each individual cloud, thus to remotely execute commands on it.
- *Virtual machine management*: it is required that the SUNFISH platform is able to access the APIs concerning the management of virtual machines, e.g. creation and deallocation.
- *Identity management*: it is required that the SUNFISH platform is able to access the local identity provider of the cloud.

These prerequisites ensure, on the one hand, that the IWM will be able to interact and control the individual clouds and, on the other hand, to federate the different identity providers.

### 5.1.1 TENANT DEFINITION MODEL

The SUNFISH tenants are introduced in Section 2.2.1. Here, to introduce how the different tenants can be deployed on the physical infrastructure, we precisely introduce the concepts of (computing) resources *belonging* to and *owned* by a cloud member. We say that a resource *belongs* to a cloud if it is physically included in its pool of resources. Instead, we say that a cloud has the *ownership* of a resource if it has exclusive access to it.

Therefore, we can characterise the different types of SUNFISH tenants as follows.

- *Infrastructure tenant*: it is formed by resources belonging to and owned by every cloud participating in the federation.
- *Standard operational tenant*: it is formed by computing resources belonging to possibly more than one member cloud and owned by one single member. Notice that when a SMC service is deployed, as it requires at least three different services, the resources must belong to more than two members.
- *Segregated operational tenant*: the tenant is formed by computing resources belonging to and owned by a single cloud member.

On the base of this tenant characterisation, we can now introduce how a tenant is effectively mapped to parts of physical resources. Therefore, we introduce the notion of *Section*, i.e. an atomic container of physical resources that are owned by a single cloud.

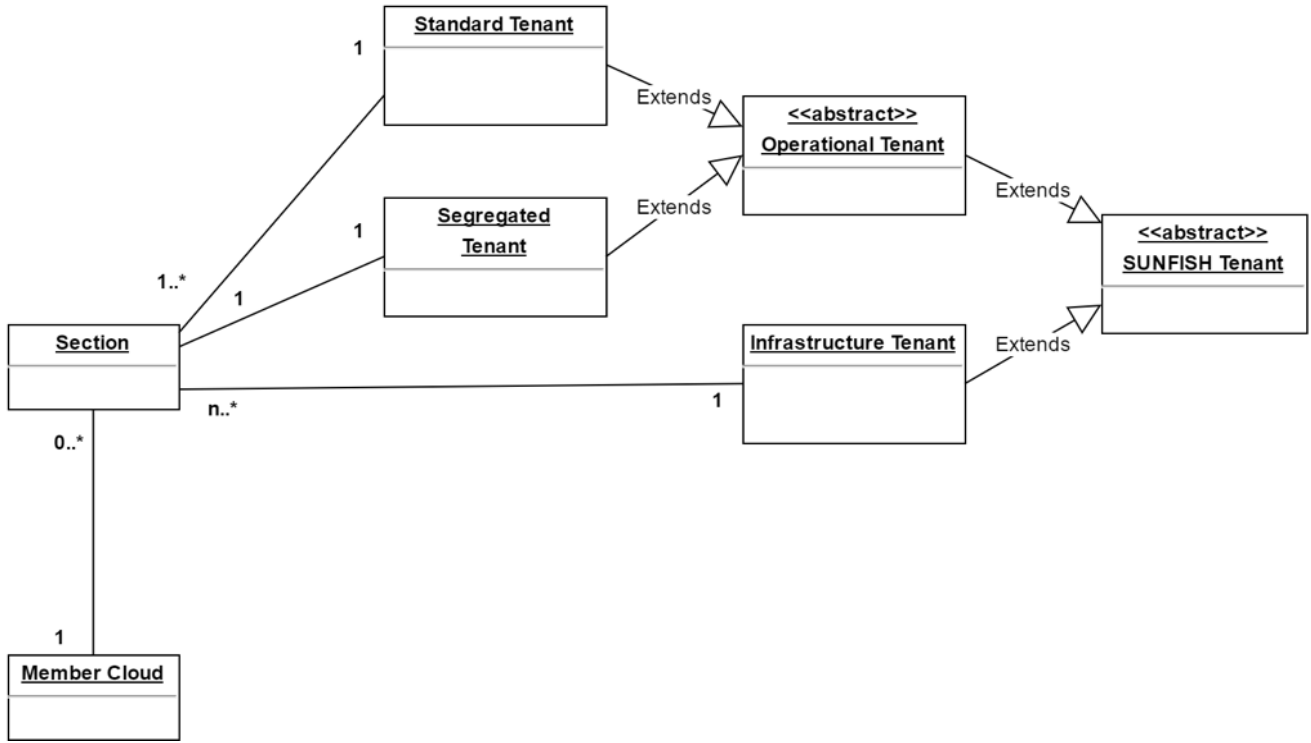

**Figure 10 Structural model of the tenants forming a FaaS federation**

A section represents the unit of physical resources that each member cloud can use to participate to a FaaS federation. Aggregating different sections permits creating containers of resources physically owned by one or more clouds and, consequently, it permits creating the various tenants. The relationships between sections and tenants are graphically represented by the model in Figure 11. The model outlines the different cardinality relationships between the tenants. Specifically, only a single section composes a segregated operational tenant (hence, it has only one owner), while multiple sections can compose a standard operational tenant (hence, it may have multiple owners). Instead, the infrastructure tenant is composed by at least $n$ sections, where $n$ is the number of federation members.

The structural model highlights how a single member cloud can contribute with its sections to the various tenants. The sections of different clouds can then be aggregated to actually create a federation. Figure 12 intuitively represents the different level of aggregation. Indeed, on the top of the physical resources of three different clouds, an infrastructural tenant and two operational tenants are created. Each circle is a section provided by a cloud to the federation. Hence, it follows from the colours that (i) each member cloud participates to the infrastructural tenant; (ii) the segregated tenant contains sections of a single cloud; (iii) the operational tenant contains sections from multiple clouds.

The creation and deployment of the tenants, according to the presented definition model, can be automated by using appropriate orchestrations of the APIs offered by the different clouds. We present in Section 5.2 a high-level view of one of these orchestrations. Before, in the following section, we outline the communications intra- and inter-tenant are performed.

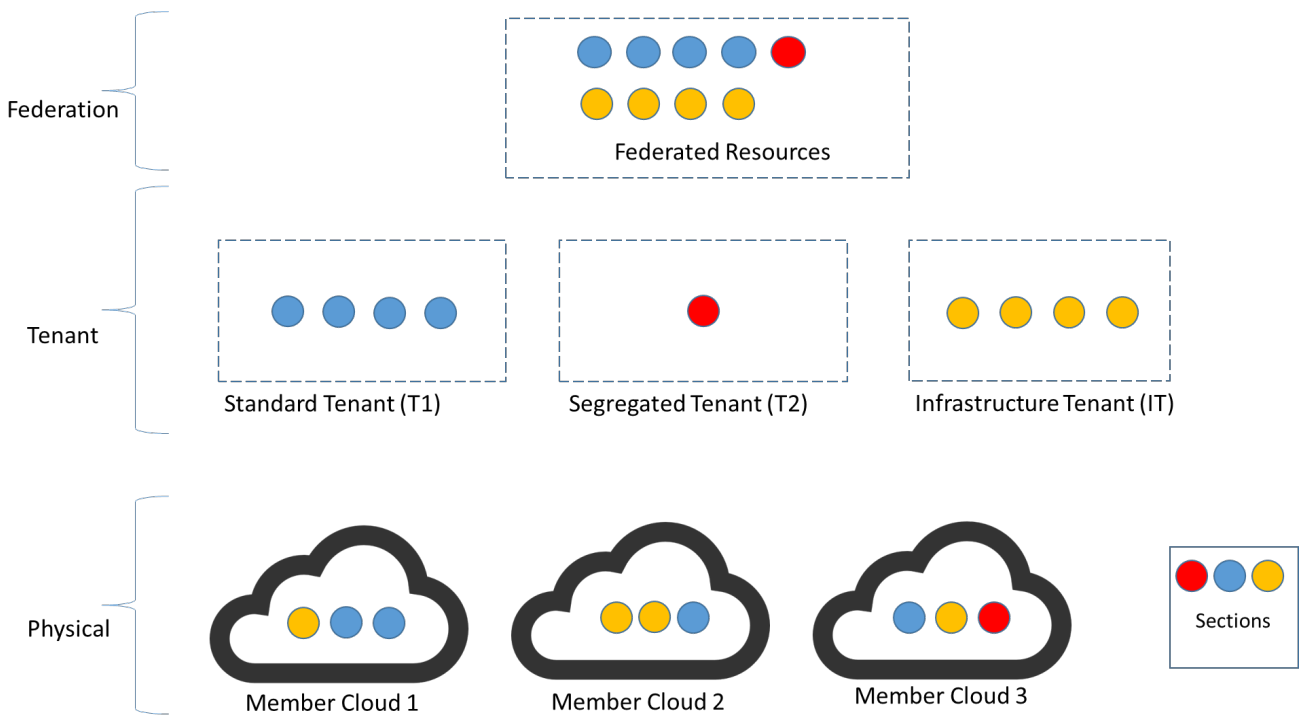

**Figure 11 Aggregation strategy of the member cloud resources**

## 5.1.2 TENANT COMMUNICATIONS

The communication between tenants is a crucial aspect to address, because, on the one hand, we need to ensure that only the appropriate endpoints are actually reachable by users outside a federation. But, on the other hand, all tenants need to communicate in order to offer the FaaS functionalities.

The basic ingredient of the communication is the tenant *visibility*, i.e. the capacity of seeing and hence communicating with tenants. The visibility among tenants is ensured by relying on a set of *infrastructure services* that are set up once a cloud federates itself to a FaaS federation. This set of services are classified as follows

- *Core Infrastructure Services:* the set of services, shared among all the federation members, realizing the core functionalities of a SUNFISH federation (i.e. enforcing inter-tenant access to services, discovering of the federated services, section management).
- *Network Infrastructure Services*: a set of services, deployed among resources belonging to a single cloud member, capable of managing at the network layer (e.g. routing and VPN) the subset of resources instantiated for the federation.
- *Access Infrastructure Services*: the set of services enabling monitored accesses (i.e., interactions) among tenants. These services are deployed in each created tenant.

These services are indeed physically distributed among the tenants, thus to enable the federation at the infrastructural level fostered by FaaS. Figure 13 reports an example of the distribution of these services. In the following, we report a few additional details on the intra- and inter-tenant communications.

The *intra-tenant* communication, i.e. between different sections of a tenant, is realized by one or more private channel created as VPN tunnels. The topology of the channels is capable of merging the various networks belonging to the sections into a single virtual network. The channels are dynamically established and maintained by means of the Network Infrastructure Services.

The *inter-tenant* communication, i.e. between different tenants, is realized by means of a bus facility shared among the whole federation. The endpoints of this bus are placed as part of the Access Infrastructure Services. This approach enables the monitoring and logging of every tenant access.

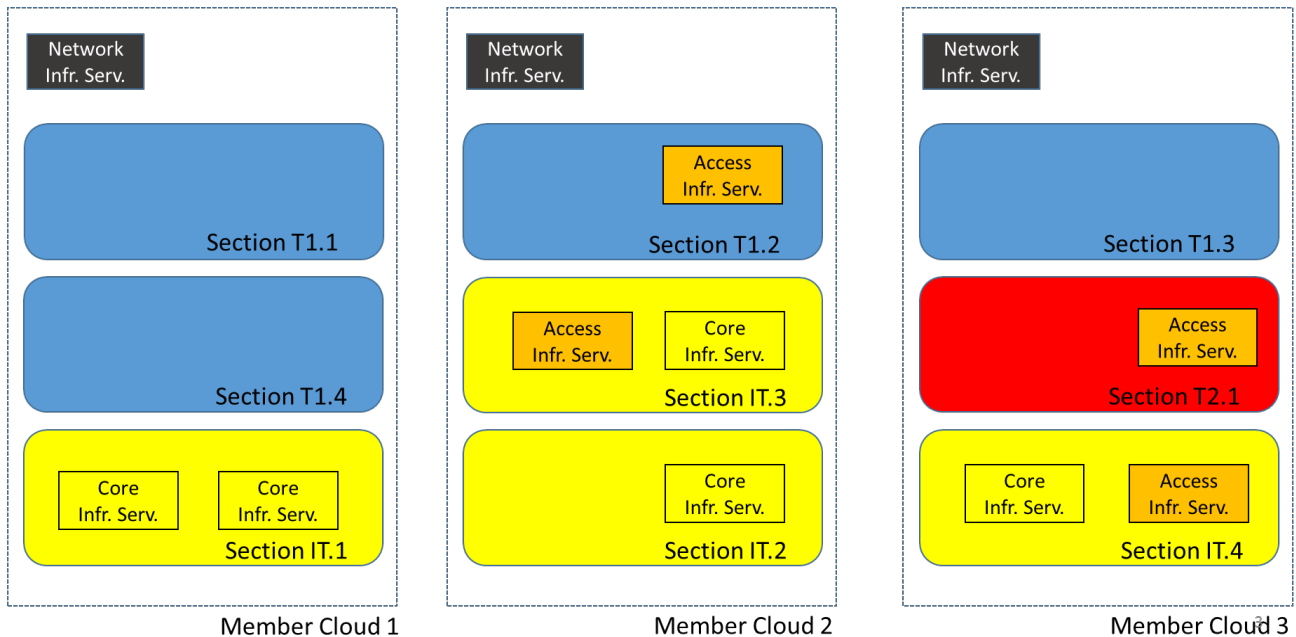

**Figure 12 Physical localisation of the infrastructure services among sections**

## 5.2 Automated Tenant Creation and Deployment

The tenant definition just presented, together with the various infrastructure services, can be automatically created and deployed by using some well-established technologies. In particular, we rely on the following ones.

- Configuration Management Engine (CME): it represents a software responsible of ensuring that a remote operating system is configured according to specific parameters. A well-established example of CME is Saltstack[5].
- Container: it represents an isolated, resource-controlled, portable operating environment where an application can be deployed. A well-established example of container is Docker[6].
- Container Cluster Manager (CCM): it represents a software capable of orchestrating Containers among a set of nodes. A well-established CCM is Kubernetes[7]

The pursued orchestrations of these technologies will be precisely presented in [16]. Here, we report the main idea behind these orchestrations related to the infrastructure tenant (Section 5.2.1) and the operational tenant (Section 5.2.2). In the former case, by way of example, we detail one orchestration.

---

[5] Saltstack - https://saltstack.com/ [17]

[6] Docker - https://www.docker.com/

[7] Kubernetes - http://kubernetes.io/

### 5.2.1 INFRASTRUCTURE TENANT

The infrastructure tenant is created once a FaaS federation is established. Then, when a new cloud joins the federation, the tenant has to be appropriately updated. We detail these two cases in the following two sections, respectively.

#### 5.2.1.1 Creation of the Tenant

The creation of the infrastructure tenant is a process started by one of the clouds willing to create the federation and, consequently, to be integrated with the other initial clouds.

The creation of the tenant starts by defining an initial section which will contain, by means of an appropriate use of a CME and a CCM, the set of Core Infrastructure Services needed to govern the federation. A similar process is performed by all the other initial clouds and hence all members will have its own section in the infrastructural tenant.

The creation of this tenant allows the cloud members to register additional sections to the federation and create new tenants.

#### 5.2.1.2 Integrating of a New Cloud in the Tenant

The integration of a new cloud "X" in an already created infrastructure tenant amounts to a precise orchestration that is carried out by an entity called Deployment Manager. This entity corresponds to a standalone software that is in charge of determining, according to the available resources, the infrastructure services to be deployed in the federation. This activity is supported by the Configurator, a static software converting higher level commands and controls (provided by the Deployment Manager) into lower level configuration activities. The Deployment manager and the Configurator are logically placed as part of the FAM, but they refer to standalone low-level software entities needed for interacting and managing different cloud systems; further details will be in [16].

This integration process occurs as part of the FaaS operating phase 1 and, more specifically, at the step 6, named "6.setupInfrFederation", of the sequence diagram in Figure 6. The corresponding orchestration corresponds to the steps reported in the sequence diagram of Figure 14. The steps are commented as follows.

- Step 6.1: the configurator interacts with the Network Infrastructure Service of the new cloud, thus to establish a direct communication for performing the deploying of the needed services.
- Step 6.2-6.3: the configurator establishes a connection with the sections that will become the tenant "T". For the sake of clarity, the term tenant "T" is reported in the diagram since the initial steps.
- Steps 6.4-6.5: the configurator collects the information about the sections to enrol to the federation (e.g., amount of offered resources) and asks to the deployment manager (i.e., a standalone software able to instantiate, according to the given values, the commands of the configurator) the services to deploy.
- Step 6.6: the configurator sends the received actions to each of the sections accordingly.
- Steps 6.7-6.8: the cloud downloads in its own turn the necessary containers (i.e., a SUNFISH-dedicated container hub will be made available). Then, it configures itself, i.e. the "Config" box in the diagram, by instantiating the requested Access Infrastructure Services. As result of this configuration, an acknowledgment is sent to the configurator.
- Steps 6.9-6.10: the configurator finalises the configuration by sending some post-configuration commands to the Network Infrastructure Service, thus to remove the communication settings only needed for this installation phase. The deployment manager is then informed of the finalisation of the process.

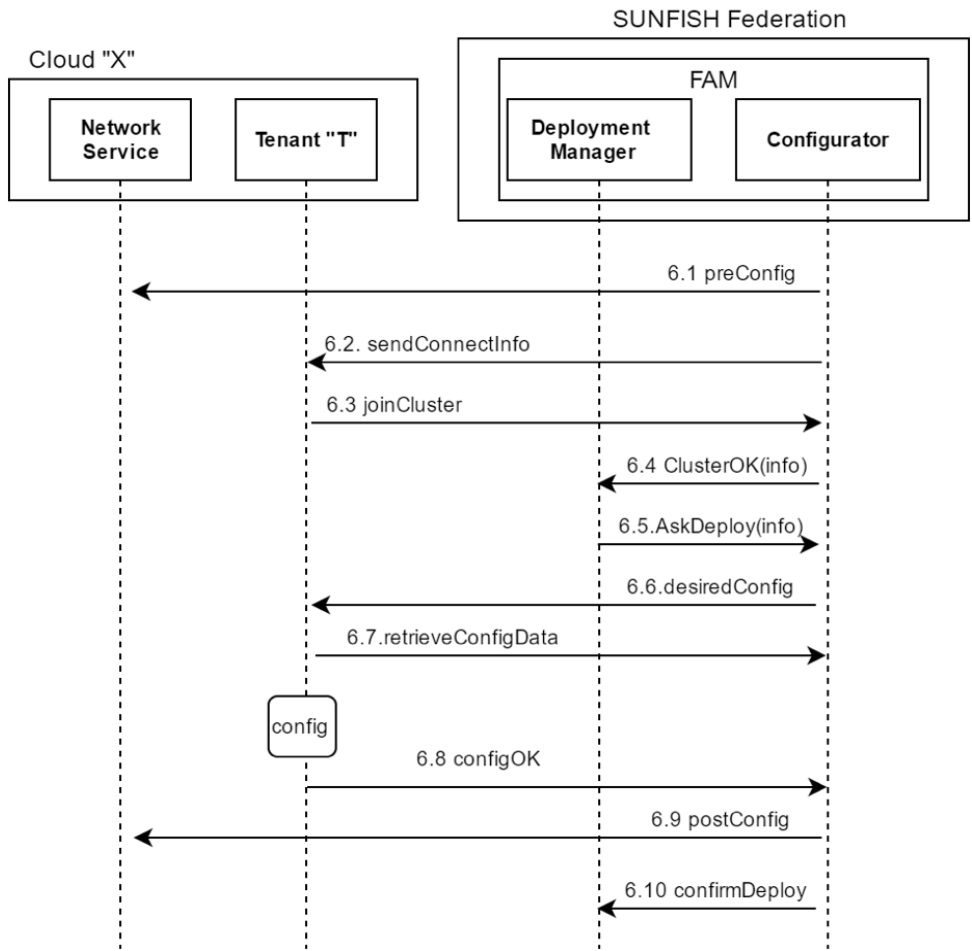

**Figure 13 Sequence diagram for the configuration of the infrastructural tenant**

## 5.2.2 OPERATIONAL TENANT

The creation of the operational tenant is realized by one single member cloud by relying on the already existing infrastructural tenant.

The creation process is similar to that of the infrastructure tenant, but a different set services and sections are used. In particular, the tenant will contain the Access Infrastructure Service properly configured, but it does not contain the Core Infrastructure Services (that are committed to only the infrastructural tenant). As previously reported, the standard and segregated tenant will differ only on the sections that will be part of the tenant.

It is finally worth noticing that the configuration and usage of the different sections is completely transparent for the cloud members.

# 6 CONCLUDING REMARKS

In this document, we have introduced and fully-formalised FaaS, the cloud federation service proposed by SUNFISH. The innovations proposed by SUNFISH consists in the conceptualisation of FaaS and its functionalities, and in the pioneering of an innovative blockchain-based realisation of the federation governance.

More specifically, we start by illustrating the basic functionalities of FaaS and its innovative features of distributed and democratic governance based on blockchain. Then, we precisely present the objectives and the SUNFISH platform that permits achieving these objectives. The platform is presented by first detailing all the functionalities of the forming components (which are summarised in Appendix B ), then by presenting their main interactions. The implementation strategies for the platform are here outlined.

# A SUNFISH TERMINOLOGY

**Federation-as-a-Service (FaaS).** The cloud federation service devised and designed by the SUNFISH project via the so-called SUNFISH Platform.

**Federated service.** It is a service offered by an individual cloud part of a FaaS federation to the other federated clouds. The service can be of any nature, from computational resources to software applications.

**Member Cloud Administrator.** It is the role corresponding to the identity assigned to the user that owns the capabilities for administering its cloud, e.g. it can create virtual machines, access all the cloud APIs, or set Service Level Agreement (SLA) and access control for the whole cloud.

**Segregated Environment.** It is a trusted environment where sensitive data can be managed with a high level of confidence. It is generated by an SUNFISH segregated operation tenant.

**Service consumer.** It is the active part requesting a federated service, i.e. either a user or a cloud application.

**Service provider.** It is the passive part providing when requested a federated service.

**SUNFISH Federation Agreement Contract (SFAC).** The contract signed by the clouds member of a FaaS federation. It contains the business goals, duties and rights of the clouds participating to the federation.

**SUNFISH Infrastructural Tenant.** It represents the backbone of the FaaS federation, as it enables, on the top of it, the main functionalities underlying FaaS. There exists a single SUNFISH infrastructure tenant for each FaaS federation.

**SUNFISH Operational Tenant.** It represents the provider of a service offered by a federation member. This type of tenant can be created in different forms according to the ownership of the used computing resources:
- *standard:* the tenant is formed by computing resources owned by different members;
- *segregated:* the tenant is formed by computing resources that are all owned by a single member.

**SUNFISH Platform.** It is the software architecture that permits realising a FaaS federation.

**SUNFISH Tenant.** It is a virtual space that is formed by resources belonging to different clouds, and can be considered as the basic building block for the federation of clouds and services

**Tenant.** It is a virtual space containing computing resources exclusively assigned to a member of a (federated) cloud.

**Tenant Administrator.** It is the role corresponding to the identity assigned to the user that owns the capabilities for administering a single tenant where a service is installed, e.g. publish new service, manage virtual networks, or set SLA and access control for a specific service provided by the tenant (in compliance with the constraint posed by the Member Cloud Administrator).

# B  SUNFISH FUNCTIONALITIES

| Federated Administration and Monitoring (FAM) |
|---|
| FAM 1 Define graphical entry-points for cloud federation administration functionalities |
| FAM 2 Report on SLA policy violation |
| FAM 3 Collect and diffuse security alerts received by the FRM and FSA components |
| FAM 4 Visualise security alerts by means of a dashboard |
| FAM 5 Subscription to federated services for service consumer |

| Identity Management (IDM) |
|---|
| IDM 1 Provide an identity to any service consumer and provider |
| IDM 2 Define a SSO authentication mechanism |
| IDM 3 Enable the federation of the identity managers of the individual clouds |
| IDM 4 Provide endpoints for the generation of authenticating crypto-tokens |
| IDM 5 Define an identity management compliant with eIDAS |

| Registry Interface (RI) |
|---|
| RI 1 Offer a set of APIs to store and retrieve the governance data to and from the blockchain-based registry |
| RI 2 Define authorisation controls on the API invocation based on crypto-tokens |

| Data Security (DS) |
|---|
| DS 1 Support the evaluation and distributed enforcement of ABAC policies |
| DS 2 Define invocation mechanisms for the DTS components |
| DS 3 Define access controls for the operational phase "Service Request" |
| DS 4 Define access controls for the operational phase "Service Usage" |
| DS 5 Define administrative controls on the modification actions on access control policies |

| Secure Multi-Party Computation (SMC) |
|---|
| SMC 1 Provide the secure SMC service |
| SMC 2 Integrate the secure SMC service with the SUNFISH platform |

### Data Masking (DM)

DM 1 Provide the processes for masking and unmasking of personal and sensitive data

DM 2 Integrate the (un)masking processes with the SUNFISH platform

DM 3 Manage the masking service state (tokenization table) via the RI

### Anonymization (ANM)

ANM 1 Provide the anonymization processes of sensitive data

ANM 2 Integrate the anonymization process with the SUNFISH platform

### Intelligent Workload Management (IWM)

IWM 1 Provide the computation of optimised federation-based workload deployment targets

IWM 2 Support different optimisation parameters for the calculation of the workload deployment targets

IWM 3 Provide support for interaction with APIs of the federated clouds, both private and public

IWM 4 Integrate DS policies with the optimisation path to filter out requestor-specific targets

### Federated Runtime Monitoring (FRM)

FRM 1 Provide distributed probes to monitor the access control system

FRM 2 Detect access control policy violations by analysing the collected access control data

FRM 3 Rise alerts to the FAM to signal access control violations

FRM 4 Provide to the FSA the access logs to perform its reasoning tasks

### Federated Security Audit (FSA)

FSA 1 Detect vulnerabilities in existing access control mechanisms by analysing access logs

FSA 2 Detect security breaches by analysing access logs

FSA 3 Rise alerts to the FAM to signal vulnerabilities and security breaches